
\def\AP#1{{\sl Ann.\ Phys.\ (N.Y.) {\bf #1}}}

\def\CMP#1{{\sl Comm. Math. Phys. {\bf #1}}}
\def\IMPA#1{{\sl Int. J. Mod. Phys. {\bf A#1}}}

\def\JMP#1{{\sl J.\ Math. \ Phys.\ {\bf #1}}}

\def\NCA#1{{\sl Nuovo Cimento {\bf #1A}}}

\def\NPB#1{{\sl Nucl.\ Phys.\ {\bf B#1}}}

\def\PLB#1{{\sl Phys.\ Lett.\ {\bf #1B}}}

\def\PRD#1{{\sl Phys.\ Rev.\ D\  {\bf #1}}}
\def\PRL#1{{\sl Phys.\ Rev.\ Lett.\ {\bf #1}}}

\def\PTP#1{{\sl Prog.\ Theor.\ Phys.\ {\bf #1}}}

\def\RMP#1{{\sl Rev.\ Mod.\ Phys.\ {\bf #1}}}

\def\TMP#1{{\sl Theor.\ Math.\ Phys.\ {\bf #1}}}


%
\def\nxl{\hfill\break}

\def\A{{\cal A}}                            
\def\B{{\cal B}}
\def\D{{\cal D}}
\def\F{{\cal F}}                            
\def\G{{\cal G}}                            
\def\H{{\cal H}}                            
\def\C{{\cal C}}                            
\def\Z{{\cal Z}}                            
\def\L{{\cal L}}                            

\def\a{\alpha}

\def\b{\beta}
\def\g{\gamma}
\def\l{\lambda}

\def\e{\epsilon}
\def\o{\over}

\def\s{\sigma}


\def\frac#1#2{{\textstyle{
 #1 \over #2 }}}                            


\def\1{{\rm 1 \!\!\, l}}                        
%

%
%


\hyphenation{Di-par-ti-men-to}
\hyphenation{na-me-ly}
\hyphenation{al-go-ri-thm}
\hyphenation{pre-ci-sion}
\hyphenation{cal-cu-la-ted}

 \input phyzzx
\def\Im{{\rm Im\,}}

 \Pubnum={$\rm PAR\; LPTHE\; 92/50;\;  hep-th / 9302052$}
 \date={}
 \titlepage
 \title{STRINGS IN CURVED SPACETIMES}
 \author{ H.J. de Vega }
 \address{ Laboratoire de Physique Th\'eorique et Hautes Energies
      \foot{Laboratoire Associ\'e au CNRS UA 280}, Paris
      \foot{Postal address: \nxl
       L.P.T.H.E., Tour 16, $1^{\rm er}$ \'etage, Universit\'e Paris VI,\nxl
	  4, Place Jussieu, 75252, Paris cedex 05, FRANCE. }}

 \author{Lectures delivered at the Erice School ``STRING QUANTUM
 GRAVITY AND PHYSICS AT THE PLANCK ENERGY SCALE'', 21-28 June 1992
, to appear in the Proceedings edited by  N. S\'anchez, World Scientific}

 \endpage
 \Pubnum{}
 \titlepage
 \title{STRINGS IN CURVED SPACETIMES}
 \author{ H.J. de Vega }
 \address{Laboratoire de Physique Th\'eorique et Hautes Energies
      , Tour 16, $1^{\rm er}$\'etage,$~~$  Universit\'e Paris VI,
	  4, Place Jussieu, 75252, Paris cedex 05, FRANCE. }

 \abstract
Progress on the physics of strings in curved spacetime are
comprehensively reviewed.
 We start by showing through renormalization group arguments that
 a meaningful quantum theory
 of gravity must be finite and must include all particle physics.
 Then, we review classical and quantum string propagation
 in curved spacetimes.We start by the general expansion method
 proposed by de Vega and S\'anchez in 1987.
The particle transmutation phenomena in
 asymptotically flat spacetimes are detailed including fermion-boson
 transitions in supergravity backgrounds.
 The next chapters review the exactly solvable cases of string
 propagation: shock waves, singular plane waves, conical spacetimes
 and de Sitter cosmological spacetime. The calculation of various
physical quantities like the string mass and the energy-momentum
tensor shows that classical and quantum string propagation in
shock-waves and singular plane waves is physically meaningful
and full of interesting new phenomena.
 The important phenomenom of {\bf string stretching}
 that takes place when strings fall into
 spacetime singularities and in expanding universes is analyzed.
 We conclude by reporting on strings in de Sitter spacetime, where
the string equations are integrable and reduce to the sinh-Gordon
equation and to integrable generalizations of it.
 \endpage

 \REF\wilson{See for example: K. G. Wilson, \RMP{47}, 773 (1975)
  and \nxl \RMP{55}, 583 (1983).}
 \REF\nos{H. J. de Vega and N. S\'anchez,  \PLB{197}, 320 (1987).}
 \REF\negro{ H. J. de Vega and N. S\'anchez,
	\NPB{309}, 552 and 577(1988).}
 \REF\ondch{H. J. de Vega and N. S\'anchez, \NPB{317}, 706 (1989) . \nxl
 D. Amati and K. Klim\^cik, \PLB{210} , 92 (1988)  , \nxl see
 also : ref.[\costa ].}
 \REF\costa{M. Costa
 and H. J. de Vega, \AP{211}, 223 and 235 (1991).}
 \REF\ondpl{H. J. de Vega and N. S\'anchez, \PRD{45} , 2783 (1992). \nxl
 H. J. de Vega, M. Ram\'on Medrano and N. S\'anchez,   LPTHE  Paris
  preprint  92-13.}
 \REF\desi{H. J. de Vega and N. S\'anchez,  LPTHE  Paris  preprint
  92-31, to appear in \PRD{}, H. J. de Vega A. V. Mikhailov and N. S\'anchez,
 LPTHE  Paris  preprint 92-32 to appear in \TMP{}, special volume in the memory
of
 M. C. Polivanov }
 \REF\vene{N. S\'anchez and G. Veneziano, \NPB{333}, 253 (1990), \nxl
 M. Gasperini, N.S\'anchez and G. Veneziano, \nxl
 \IMPA{6}, 3853 (1991) and \NPB{364}, 365 (1991).}
 \REF\tanos{ M. Ademollo, A. D'Adda, R. D'Auria, E. Napolitano, S. Sciuto,
	 P. Di Vecchia, F. Gliozzi, R. Musto and F. Nicodemi,
	 \NCA{21}, 77(1974). }
 \REF\prim{ H. J. de Vega and N. S\'anchez, \NPB{299}, 818(1988).}
 \REF\conico{H. J. de Vega and N. S\'anchez, \PRD{42} , 3969 (1992) and \nxl
 H. J. de Vega, M. Ram\'on Medrano and N. S\'anchez,\NPB{374}, 405 (1992)}
 \REF\bdvhp{ L. Brink, P. Di Vecchia and P. Howe,
	 \PLB{65}, 471 (1976).\nxl
	 A.M. Polyakov, \PLB {103}, 207 and 211(1981).}
 \REF\japo{ T. Goto, \PTP{46}, 1560(1971).
	 O. Hara, \PTP{46}, 1549(1971).
	 Y. Nambu, Lectures at the Copenhagen Symposium.}
 \REF\gsws{ See for example: M. Green, J. S. Schwarz and E. Witten,
	 Superstrings vol. I, \nxl
	 J. Scherk, \RMP{47}, 123(1975).}
 \REF\plb{H. J. de Vega and N. S\'anchez, \PLB {244}, 215(1990).\nxl
    H. J. de Vega and N. S\'anchez,
	 \PRL{65} (C), 1517(1990).}
 \REF\trans{ H. J. de Vega, M. Ram\'on Medrano and N. S\'anchez,
 \NPB{351}, 277 (1991).}
 \REF\stran{H. J. de Vega, M. Ram\'on Medrano and N. S\'anchez,
 \NPB{374}, 425 (1992).}
 \REF\super{H.J.de Vega, M.Ram\'on Medrano and N. S\'anchez,
\PLB{285},206(1992).}
 \REF\camas{H. J. de Vega and N. S\'anchez, \NPB{317}, 731 (1989).}
 \REF\eneimp{H. J. de Vega and N. S\'anchez, \IMPA{7}, 3043 (1992).}
 \REF\horo{ G. Horowitz and A.R. Steif, \PRL{64}, 260(1990)and \nxl
 \PRD{42} , 1950 (1990).}
 \REF\gth{ G. 't Hooft, \PLB{198}, 61 (1987).}
 \REF\bida{ See for example, N. D. Birrell and P. C. W. Davies,
 Quantum Fields in curved spaces,Cambridge University Press, 1982 }
 \REF\guven{R. G\"uven, \PLB{191}, 275 (1987).}
\REF\pol{K. Pohlmeyer \CMP{46}, 207 (1976) \nxl
         H. Eichenherr and J. Honerkamp, \JMP{22}, 374 (1981).}

  \chapter{Quantum Gravitation at the light of the Renormalization Group.}

	 The construction of a sensible quantum theory of gravitation is
 probably the greatest challenge in today's theoretical physics.
	 Deep problems arise when (second) quantization is combined with
 general relativity. Statistical phenomena show up (Hawking's radiation)
 when free fields are quantized in black-holes backgrounds. This entails a
 lack of quantum coherence even keeping the gravitational field classical.

	 Another problem (the most often discussed in this connection)
  is the one of renormalizability of the Einstein theory
 (or its various generalizations) when quantized as a local quantum field
 theory. Actually, even deeper conceptual problems arise when one tries to
 combine quantum concepts with General Relativity. That is, it may be very
 well that a quantum theory of gravitation needs new concepts and ideas.
 Of course, this future theory must have the today's General Relativity and
 Quantum Mechanics (and QFT) as limiting cases. In some sense, what everybody
  is doing in this domain (including string theories approach) may be
  to the real
 theory what the old quantum theory in the 10's was compared with
 quantum mechanics.

	 The main drawback to develop a quantum theory of gravitation is
 clearly the total lack of experimental guides for the theoretical
 developpment. Just by dimensional reasons, physical effects combining
 gravitation and quantum mechanics are relevant only at energies of the
 order of  $M_{Planck}  =  \hbar c / G  =  1.22 \,  10^{16} $Tev.
 Such energies were available in the Universe at times $ t < t_{Planck}
 =  5.4 \, 10^{-44} $sec. Anyway, as a question of principle,
 to construct a quantum theory of gravitation is a problem of fundamental
  relevance for theoretical physics . In addition, one cannot rule
 out completely the possibility of some ``low energy'' ($E \ll  M_{Planck}$)
 physical effect that could be experimentally tested.

	 Let us discuss now from a conceptual point of view the
  renormalizability question for gravitation. What is a renormalizable QFT?
 It is a theory with a domain of validity characterized by energies E
 such that [\wilson]
 $$
 E  < \Lambda
 $$
 Here, the scale  $\Lambda $  is proper to the model under consideration (e. g.
 , $\Lambda \simeq 1 $ GeV for QED, $ \Lambda \simeq  100 $ GeV for the
 standard model of strong and electroweak  interactions, etc.).
 One always applies the QFT in question till infinite energy (or zero distance)
  for virtual processes and finds usually ultra-violet infinities.
 These divergences reflect the fact that the model becomes unphysical
 for energies $ \Lambda \ll  E \leq \infty $. In a renormalizable QFT these
 infinities can be absorbed in a finite number (usually a few) parameters
 like coupling constants and mass ratios. Since these parameters
 (usually called renormalized masses and couplings) are not predicted
 by the model in question, one has to fit them to their experimental values.
 One needs a more general theory valid at energies beyond  $\Lambda$
   in order to compute these renormalized parameters (presumably from
 more fundamental constants). For example,  $M_W/M_Z $  is calculable in
 a GUT whereas it must be fitted to its experimental value in the standard
 electroweak model.

	 Let us now see what are the consequences of Heisenberg's principle
 in quantum mechanics combined with the notion of gravitational
 (Schwarzschild) radius in General Relativity.
	 Assume we make two measurements at a very small distance $\Delta x$ .
  Then,
 $$
 \Delta p  \sim \Delta E  \sim  1 / \Delta x
 $$
 where we set  $\hbar = c = 1$ . For sufficiently large  $\Delta E$,
 particles with masses $m \sim  1/\Delta x$ will be produced.
 The gravitational radius of such particles are of the order
 $$
		 R_G \sim G m \sim {(l_{Planck})^2 \o {\Delta x}}
	      \eqn\heis
 $$
 where  $l_{Planck}  \sim 10^{-33}$ cm. Now, General Relativity allows
 measures at a distance  $\Delta x$ , provided
 $$
  \Delta x  > R_G  \quad \to \quad  \Delta x   >  {(l_{Planck})^2 \o{ \Delta
x}}\qquad .
 $$
 That is,
 $$
		  \Delta x  > l_{Planck} \quad or \quad  m   <   M_{Planck}
 \eqn\colap
 $$
 This means that no measurements can be made at distances smaller than the
 Planck length and that no particle can be heavier than  $M_{Planck}$ .
  This is a simple consequence of relativistic quantum mechanics combined
  with General Relativity. In addition, the notion of locality and hence
 of spacetime becomes meaningless at the Planck scale. Notice that the
 equality in eq.\colap\ corresponds when the Compton length equals
 the Schwarzschild radius of a particle.
   Since  $M_{Planck}$ is the heaviest possible particle scale, a theory valid
 there (necessarily involving quantum gravitation) will also be valid at any
lower
 energy scale. One may ignore higher energy phenomena in a low energy theory,
  but not the opposite. In other words, it will be a `theory of everything'.
 We think that this is the {\bf key point} on the quantization of gravity.
 A theory that holds till the Planck scale must describe {\bf all} what happens
 at lower energies including all known particle physics as well as what
  we do not know yet (that is, beyond the standard model).
 Notice that this conclusion is totally
 independent of the use of string models.
  A direct important consequence of this conclusion, is that it does not
  make physical sense to quantize {\bf pure gravity}. A physically
 sensible quantum theory cannot contain only gravitons. To give an example,
  a theoretical prediction for graviton-graviton scattering at energies of
 the order of $M_{Planck}$   must include all particles produced in a real
 experiment. That is, in practice, all existing particles in nature, since
 gravity couples to all matter.

	 Let us now come to the renormalizability problem for gravitation.
 As is clear from the preceding discussion, we have  $\Lambda  \simeq
  M_{Planck}$ for gravitation. There cannot be any theory of
 particles beyond it. Therefore, if ultraviolet divergences appear in a
 quantum theory of gravitation, there is no way to interpret them as coming
 from a higher energy scale as it is usually done in QFT (see above).
  That is, no physical understanding can be given to such UV infinities.
 The only logically consistent possibility would be to find a {\bf finite
 theory} of quantum gravitation which is a TOE.

	 These simple arguments, based on the renormalization group lead us to
  an important conclusion : a consistent quantum theory of gravitation must be
  a finite theory and must include all other interactions. That is, it must be
  a theory of everything (TOE). This is a very ambitious project. In particular
  it needs the understanding of the present desert between 1 and $10^{16}$ TeV.
	 There is an additional dimensional argument about the inference
 Quantum Theory of Gravitation $\to$  TOE. There are only three dimensional
 physical magnitudes in nature: length, energy and time and correspondingly
 only three dimensional constants in nature: c, h and G. All other physical
 constants like  $\a  = 1/137,04... , M_{proton}/m_{electron}, \, \theta_{WS}$
 ,... etc. are pure numbers and they must be calculable in a TOE.
 This is a formidable, but extremely appealing problem.
	 From the theoretical side, the only serious candidate for a TOE
  is at present string theory. This is why we think that strings desserve a
 special attention in order to quantize gravity.

	 As a first step on the understanding of quantum gravitational
 phenomena in a string framework, we started in 1987 a programme of string
 quantization on curved spacetimes [\nos,\negro]. The investigation of
strings in curved spacetimes is currently the best framework to study the
physics of gravitation in the context of string theory, in spite of
its limitations. First, the use of a continuous Riemanian manifold to
describe the spacetime cannot be valid at scales of the order of $l_{Planck}$.
More important, gravitational backgrounds effectively provide classical
or semiclasical descriptions even if the matter backreaction to the
geometry is included through semiclassical Einstein equations (or
stringy corrected Einstein equations) by inserting the expectation
value of the string energy-momentum tensor in the r.h.s. One would
want a full quantum treatment for matter and geometry. However,to find
a formulation  of string theory going beyond the use of classical backgrounds
 is a very difficult (but  fundamental!) problem.  One would like
 to derive the spacetime geometry as a classical and low energy
  ($ \ll M_{Planck}$) limit from the solution of (quantum) string theory.

 \chapter{ Strings in Curved Spacetimes. Introduction.}

	 Let us consider bosonic strings (open or closed) propagating in a curved
D-dimensional spacetime defined by a metric $G_{AB}(X), 0 \leq A,B \leq D-1$.
  The action can be written as
 $$
    S  = {{1}\o{2 \pi \a'}} \int d\s d\tau \sqrt{g}  g_{\a\b}(\s,\tau) \,
 G_{AB}(X) \,
 \partial^{\a}X^A(\s,\tau) \, \partial^{\b}X^B(\s,\tau)
 \eqn\accion
 $$
 Here  $ g_{\a\b}(\s,\tau)$  ( $0 \leq \a, \b \leq 1$ ) is the metric in the
 worldsheet, $\a'$ stands for the string tension. As in flat spacetime, $\a'
 \sim (M_{Planck})^{-2} \sim ( l_{Planck})^2$  fixes the scale in the theory.
  There are no other free parameters like coupling constants in string theory.
  Besides eq. \accion\ which is the curved spacetime version of the
 Brink-DiVecchia-Howe-Polyakov action [\bdvhp] , one can also start from the
 Goto-Nambu action [\japo ]  which is classically equivalent to \accion\ .

 The string action \accion\ classically enjoys Weyl invariance
 on the world sheet
 $$
  g_{\a\b}(\s,\tau) \to \l(\s,\tau) \,  g_{\a\b}(\s,\tau)
 \eqn\weyl
 $$
 plus the reparametrization invariance
 $$
 \s \to  \s'  =  f(\s,\tau) \qquad , \qquad \tau \to \tau'   =  g(\s,\tau)
	 \eqn\reparam
 $$
	 The dynamical variables being here the string coordinates
  $X_A(\s,\tau)$ , ($0 \leq A \leq D-1$) and the world-sheet metric
 $ g_{\a\b}(\s,\tau)$  .
  Extremizing  $S$  with respect to them yields the classical equations of
motion:
 $$\eqalign{
 \partial^{\a}[\sqrt g\, G_{AB}(X) \partial_{\a}X^{B}(\s,\tau)]  =
 (1/2) \sqrt g\, \partial_{A}G_{CD}(X) \, & \partial_{\a}X^{C}(\s,\tau)
  \, \partial^{\a}X^{D}(\s,\tau)\cr
  & 0 \le A \le D-1 \cr}
 \eqn\movi
 $$
 $$\eqalign{
 T_{\a\b} ~ \equiv & ~ G_{AB}(X)[ \, \partial_{\a}X^{A}(\s,\tau) \,
  \partial_{\b}X^{B}(\s,\tau) \cr
   - & (1/2) \, g_{\a\b}(\s,\tau) \,  \partial_{\g}X^{A}(\s,\tau) \,
 \partial^{\g}X^{B}(\s,\tau)\, ]= 0 ~~,
  \quad 0 \le \a , \beta \le 1. \cr}
 \eqn\vincu
 $$
 Eqs.\vincu\  contain only first derivatives and are therefore a set
  of constraints. Classically, we can always use the reparametrization
 freedom \reparam\ to recast the world-sheet metric on diagonal form
 $$
	  g_{\a\b}(\s,\tau)  =  \exp[ \phi(\s,\tau) ]\, \, {\rm diag}( -1, +1 )
 $$
 In this conformal gauge, eqs.\movi\ - \vincu\ take the simpler form:
 $$\eqalign{
 \partial_{-+}X^{A}(\s,\tau)  +   \Gamma^{A}_{BC}(X)\, & \partial_{+}
  X^{B}(\s,\tau) \, \partial_{-}X^{C}(\s,\tau) =  0~, \quad
   0 \le A \le D-1 , \cr}
 \eqn\conouno
 $$
 $$
	 T_{\pm\pm} \equiv G_{AB}(X) \, \partial_{\pm}X^{A}(\s,\tau) \,
  \partial_{\pm}X^{B}(\s,\tau)\equiv 0\, , \quad  T_{+-} \equiv T_{-+} \equiv 0
 \eqn\conodos
 $$
 where we introduce light-cone variables  $x_{\pm} \equiv \s \pm \tau $
   on the world-sheet and where  $\Gamma^{A}_{BC}(X)$  stand for the
 Christoffel symbols associated to the metric  $ G_{AB}(X)$ .

 The string boundary conditions in curved spacetimes are identical to
 those in Minkows- ki spacetime. That is,
 $$\eqalign{
  X^{A}(\s + 2 \pi,\tau) = \,   X^{A}(\s,\tau) \quad & closed \,strings \cr
    \partial_{\s}X^{A}(0,\tau)=\, \partial_{\s} X^{A}(\pi,\tau) =  \,0
 \quad & open \,strings \cr}
 \eqn\condc
 $$
	 In flat spacetime eqs.\conouno\  are linear and one can solve them
 explicitly as well as the quadratic constraint \conodos\ [\gsws].
 The solution of eqs.\conouno\  in Minkowski spacetime is usually written for
 closed strings as
 $$
	 X^{A}(\s,\tau)  =  q^A + 2 p^A \a' \tau + i \sqrt{\a'} \sum_{n \neq 0}
   \{ \a^{A}_{n} \exp[in(\s - \tau)]
	 + \tilde  \a^{A}_{n} \exp[-in(\s + \tau )] \}/n
 \eqn\solm
 $$
 where  $q^A$  and  $p^A$  stand for the string center of mass position
  and momentum and  $ \a^{A}_{n}$  and $\tilde \a^{A}_{n}$  describe the right
 and left oscillator modes of the string, respectively. This resolution is no
more
 possible in general for curved spacetime where the equations of motion
 \conouno\  are non-linear. In that case, right and left movers
 interact with themselves and with each other.
 For some spacetimes eqs.\conouno\ - \conodos\  are exactly
  solvable [see secs. 4-6]. In addition, the question of how to define
  particle states in curved spacetime appears since no preferred frames
 exist in General Relativity. These questions already appeared in
 field theory (for the string case, see ref.[\prim]).
	 In all our treatment we consider a given geometry  $ G_{AB}(X)$
   where our strings propagate. That is, we are dealing with test strings
  and we neglect for the moment their backreaction on the metric.
  Up to now, the string propagation has been solved in the following
 spacetimes thanks to specific features proper to each of them :

 1)Linear graviton [\tanos].

 2)Shock-wave spacetimes (Aichelburg-Sexl metrics and generalizations)
 [\ondch -\costa].

 3)Non-linear gravitational plane waves [\ondpl -\horo ].

 4)Conical spacetimes (the geometry around a cosmic string)[\conico].

 5)De Sitter cosmological spacetime [\desi].

	 Let us describe now the general scheme proposed in
ref.[\nos] to solve the string equation
  of motion and constraints both classically and quantum mechanically.
 The principle is the following, we start from an
 exact solution of eq.\movi\  and develop in perturbations around it. We set
 $$
 X^A(\s,\tau)  =  q^A(\s,\tau)  +  \eta^A(\s,\tau) +  \xi^A(\s,\tau)  + .....
\eqn\metod
 $$
 Here $q^A(\s,\tau)$  is an exact solution of eq.\movi\ and $\eta^A(\s,\tau)$
  obeys the linearized perturbation around  $q^A(\s,\tau)$ :
 $$
			 D^{A}_{B} \, \eta^B(\s,\tau)=  0
 \eqn\fluc
 $$
 where
 $$
 D^{A}_{B}   =  \delta^{A}_{B} \partial^2   +  \Gamma^{A}_{BC}(q)
 (\partial_{-}q^{C} \partial_{+} + \partial_{+}q^{C} \partial_{-} )
 +\partial_{B}\Gamma^{A}_{CD}(q) \partial_{+}q^{C} \partial_{-}q^{D}
	 \eqn\etae
 $$
 Here $\partial^2 \equiv \partial_{-+}$ and  $\xi_{A}(\s,\tau)$
  is a solution of the second order perturbation
  around  $q_{A}(\s,\tau)$
 $$\eqalign{
 D^{A}_{B}\, \xi^B(\s,\tau) =  &- \eta^D \partial_{D}\Gamma^{A}_{BC}(q)
  (\partial_{-}q^{C} \partial_{+} + \partial_{+}q^{C} \partial_{-} ) \eta^B
	 \cr & -   \Gamma^{A}_{BC}(q)  \partial_{+} \eta^B\partial_{-}\eta^C -
 (1/2) \eta^D \eta^E \partial^2_{DE} \Gamma^{A}_{BC}(q)  \partial_{+}q^{B}
 \partial_{-}q^{C} \cr}
 \eqn\tzi
 $$
 Higher order perturbations can be considered systematically, but we will
  restrict here to first and second orders. Notice that eq.\fluc\ is
 homogeneous whereas eq.\tzi\ is inhomogeneous with the r.h.s.
 quadratic in the $\eta$'s , solutions of eq.\fluc\ .

	 The choice of the starting solution is upon physical insight.
 Usually we start from the solution describing the center of mass motion
 of the string $ q^A(\tau)$ where
 $$
 \ddot q^{A}(\tau)  +   \Gamma^{A}_{BC} (q) \, \dot q^{B}(\tau)\,
 \dot q^{C}(\tau)  =  0
 \eqn\cm
 $$
 The world-sheet time variable is here identified with the proper time of
  the center of mass trajectory.
	 It must be noticed that we are treating the space-time geometry
 {\bf exactly}  and taking the string oscillations around  $ q^A(\s,\tau)$
  [for example the centre of mass solution  $ q^A(\tau)$  of eq.\cm ]
  as perturbations. So, our expansion corresponds to low energy excitations
 of the string as compared with the energy associated to the geometry.
 In a cosmological or black hole metric, our method corresponds to an
 expansion in  $\omega /M$ , where $\omega$ is the string frequency
  mode and  $M$  is the universe mass or the black hole mass respectively.
 This can be equivalently considered as an expansion in powers of
  $ \sqrt{\a'}$ . Actually, since   $\a' \sim (l_{Planck})^2$ ,
  the expansion parameter turns out to be the dimensionless constant
 $$
		 g  \equiv   l_{Planck} / R_{c} \simeq   1/( l_{Planck} M)
    \simeq   \omega/M
 \eqn\ge
 $$
 where $R_{c}$ characterizes the curvature  of the space-time under
 consideration .
 In most of the interesting situations one clearly has  $g \ll 1$.

	 The constraint equations \vincu\  must also be expanded in perturbations.
  We find up to terms of order higher than the second
 $$\eqalign{
    2\pi\a'  T_{\pm\pm}  = & \,  G_{AB}(q)\partial_{\pm} q^A \partial_{\pm}q^B
  + \eta^C \partial_{C}  G_{AB}(q) \partial_{\pm}q^A  \partial_{\pm}q^B
  \cr & +   2 \, G_{AB}(q)  \partial_{\pm}q^A \partial_{\pm} \eta^ B +
  (1/2)  \eta^C \eta^D  \partial_{CD}G_{AB}(q)  \partial_{\pm}q^A
 \partial_{\pm} q^B   \cr
   & +  G_{AB}(q) \partial_{\pm}\eta^ A  \partial_{\pm}\eta^B  +
   2G_{AB}(q) \partial_{\pm} q^A  \partial_{\pm}\xi^B    \cr
   & +  2\eta^C \partial_{C} G_{AB}(q) \partial_{\pm} q^A  \partial_{\pm}\eta^B
 +
   \xi^C  \partial_{C}G_{AB}(q)
 \partial_{\pm}q^A \partial_{\pm} q^B  =  0 \cr}
		 \eqn\pervin
 $$
 From this, we define the Virasoro generators  $L_n$  as
 $$
 T_{\pm\pm} =  {1 \o {2\pi}}  \sum_{n \e Z}  L_n  \exp[in( \s \pm \tau )]
		 \eqn\vira
 $$
	 It must be noticed that one must solve the linear equation
  \tzi\ , expressing
  the solution  $\xi^{A}(\s,\tau)$, as a bilinear functional of
  $ \eta^A(\s,\tau)$ . Then, one must insert the $\eta^A(\s,\tau)$
   and $\xi^A(\s,\tau)$
  into the constraints \pervin\  in order to obtain the $L_n$.
  In this approximation the  $L_n$  turn to be bilinear in the
   $\eta^A(\s,\tau)$.

 Let us now consider the case where the zeroth-order (exact) solution
 $q_A(\tau)$  describes the center of mass motion. In this case the
 fluctuation equations simplify to a set of ordinary differential equations
  [\nos ] . That is, setting
 $$
 \eta^A(\s,\tau) = \sum_{n \e Z} e^{in\s} g^{A}_n (\tau)
 \eqn\fou
 $$
 the $ g^{A}_n (\tau)$ satisfy a system of D coupled ordinary
 differential equations:
 $$
 \left[ \delta^{A}_{B} \, ( {{d^2}\o{d\tau^2}} + n^2) + 2\,
 \Gamma^{A}_{BC}(q)\, \dot q^{C}(\tau) +\dot q^{C}(\tau)\, \dot q^{D}(\tau)
 \, \partial_{B} \Gamma^{A}_{CD} (q)\right]g^{B}_n (\tau)=0
 \eqn\flut
 $$
 Notice that the different n-modes are decoupled.
	 If we apply our method to Minkowski spacetime, the
 zero order solution of eq.\cm\ , $q^A(\tau)$ , plus the first fluctuations
   $ \eta^A(\s,\tau)$  provide the exact solution of the string equations.

 The use of this expansion method applied to cosmological spacetimes
 showed the appearence of unstabilities
 in the case of de Sitter [\nos ,\vene ]. That is, exponentially growing modes
 $ \eta^A(\s,\tau)$ . In principle, one cannot believe anymore an expansion
 like this when the first order blows up. However, it turns out that these
 unstabilities far to be artifacts of the approximation used,
 revealed a true physical phenomenom : string stretching. That is, the string
 size grows indefinetely with time. This phenomenon, typical of strings
 falling into a singularity or in an inflationary universe is confirmed
 by the exact resolution of the string equations (see secs.4 and 6 and
ref.[\ondpl ,\desi ]).

 \chapter{Strings in asymptotically flat spacetimes. Particle Transmutation.}

	 When the metric   $G_{AB}(X)$  admits flat spacetime regions, say (+)
  and (-) ( i.e.  $ G_{AB}(X)$ $ \to ~~ \eta_{AB} $ ) , as it is
 the case for black-holes, for shock-wave spacetimes and generally
 for asymptotically flat geometries,
 then the equations of motion \conouno\  become in these regions
 $$
	 \partial^2 X^{A}(\s,\tau)_{\pm}  =  0
    \eqn\mink
 $$
 In particular,  $\partial^2 U(\s,\tau)  =  0$ (where $ U = X^0 - X^{D-1}$),
  and the light-cone
 gauge identification can be chosen
 $$
			 U  = 2 \, \a' p^U \tau
 \eqn\gauge
 $$
 This enables us to define ingoing and outgoing solutions for  $\tau \to
 - \infty$  and  $\tau \to + \infty$, respectively. That is,
 $$
  X^{A}(\s,\tau)_{\pm}  =  q^A_{\pm} + 2 p^A_{\pm} \a' \tau +
  i \sqrt{\a'} \sum_{n \neq 0}
   \{ \a^{A}_{n\pm} \exp[in(\s - \tau)]
	 + \tilde  \a^{A}_{n\pm} \exp[-in(\s + \tau )] \}/n
 \eqn\sola
 $$
 The connection between the out operators $\{\a^{A}_{n+} , \tilde
 \a^{A}_{n+} ,n \e \Z \,(n \neq 0),
 q^{A}_{-}, p^{A}_{-};0 \leq A \leq D-1\}$ and the in operators
 $\{ \a^{A}_{n-} ,  \tilde \a^{A}_{n-} , n \e \Z\,(n \neq 0),
 q^{A}_{-}, p^{A}_{-};0 \leq A \leq D-1\}$ depends
  on the interaction with the geometry in the non-flat regions.
  In general, the exact relation between (+) and (-) operators is
 highly non-trivial, involving all the in operators through a non-linear
  transformation
 $$\eqalign{
   \a^{A}_{n+} \, = & \, F_{n}^{A}( \a^{B}_{m-} , \tilde \a^{B}_{m-},
 q^{B}_{-}, p^{B}_{-}; m \e Z \,(m \neq 0), 0 \leq B \leq D-1 )\cr
 \tilde \a^{A}_{n+}  =&  \, G_{n}^{A}( \a^{B}_{m-} , \tilde \a^{B}_{m-},
 q^{B}_{-}, p^{B}_{-};
 m \e Z\,(m \neq 0), 0 \leq B \leq D-1 )\cr}
	 \eqn\trafg
 $$
 This transformation contains in principle all the information about the
  scattering of the string by the geometry. However, it is very difficult to
 obtain the functions  $F_{n}^{A}$  except for the exactly solvable
 cases (see secs.4-5).
  It is possible to make a detailed analysis in the context of the
 linearization method described in sec.II. [eqs.\fluc\ -\cm ]. In this
 approximation eq.\trafg\  becomes a linear, e.g. a Bogoliubov transformation.
 We have in this scheme,
 $$
  \partial^2 \eta^{A}(\s,\tau)_{\pm}  =  0
    \eqn\meta
 $$
 We can define a basis of solutions  $f^{AB}_{n\pm}(\tau)$  of
 eq.\flut\  by selecting their asymptotic behavior:
 $$
 \lim_{\tau\to\pm\infty}  f^{AB}_{n\pm}(\tau)  =   \delta^{AB}  \exp(-in\tau)
 \eqn\deff
 $$
 The choice of a positive frequency factor  $\exp(-in\tau)$  corresponds to in
or
 out particle states for  $\tau \to -\infty$ or  $\tau \to +\infty$
 respectively. For each value  $(A,B)$ we have two basis  (-)  and (+).
 Since we have free oscillators in both asymptotic limits  $\tau \to
\pm\infty$,
 $$
 \lim_{\tau\to\mp\infty}  f^{AB}_{n\pm}(\tau)  =   \A^{AB}_{n\pm}\exp(-in\tau)+
  \B^{AB}_{n\pm} \exp(+in\tau)
 \eqn\jost
 $$
 where the constant coefficients $\A^{AB}_{n\pm}$  and $\B^{AB}_{n\pm}$
   depend on the detailed time evolution for all  $\tau$.
	 We can expand the string fluctuations  $\eta^{A}(\s,\tau)$
  in the basis $f^{AB}_{n+}(\tau)$  and in the basis
 $f^{AB}_{n-}(\tau)$  as:
 $$\eqalign{
  \eta^{A}(\s,\tau)= &i \sqrt{\a'} \sum_{n \neq 0}\sum_{B=0}^{D-1}
 f^{AB}_{n\pm}(\tau)
   \{ \a^{B}_{n\pm} \exp\left[in\s\right]
	 + \tilde\a^{B}_{n\pm} \exp\left[-in\s\right] \}/n\cr}
 \eqn\desa
 $$
 Then, from eqs.\deff\ -\desa\ , we have
 $$\eqalign{
 \eta^{A}(\s,\tau)_{-}= &\lim_{\tau\to - \infty}\eta^{A}(\s,\tau)
 =  \cr
  & i \sqrt{\a'} \sum_{n \neq 0}
   \exp\left[-in\tau\right]\{ \a^{A}_{n-} \exp[in\s]
	 + \tilde  \a^{A}_{n-} \exp\left[-in\s \right] \}/n \cr
 \eta^{A}(\s,\tau)_{+}=& \lim_{\tau\to + \infty}\eta^{A}(\s,\tau)
 = \cr & i \sqrt{\a'} \sum_{n \neq 0}
   \exp\left[-in\tau\right]\{ \a^{A}_{n+} \exp\left[in \s \right]
	 + \tilde  \a^{A}_{n+} \exp\left[-in \s \right] \}/n \cr}
 \eqn\etamen
 $$
 and
 $$\eqalign{
 \a^{A}_{n+}= & \sum_{B=0}^{D-1}\left[ \A^{AB}_{n-}\a^{B}_{n-}-
 \B^{AB}_{n-}(\tilde\a^{B}_{n-})^{\dagger} \right] \cr
 \tilde\a^{A}_{n+}= & \sum_{B=0}^{D-1}\left[ \A^{AB}_{n-}\tilde\a^{B}_{n-}-
 \B^{AB}_{n-}(\a^{B}_{n-})^{\dagger} \right]\cr}
 \eqn\bogo
 $$
 This is a Bogoliubov transformation, giving the outgoing operators
 $\a^{B}_{n+} ,\tilde  \a^{B}_{n+}$
 as a {\bf linear} superposition of the creation and annhilation ingoing
 operators  $\a^{B}_{n-} ,\tilde  \a^{B}_{n-}$.

  We see that for fixed  $n$, transitions take place between
 the internal oscillatory modes of the string. The scattering of the
 string by the curved geometry involves two main effects:

 (i)	Polarization changes in the modes (without changing their right
 or left character).

 (ii) 	A mixing of particle and antiparticle modes, changing at the
 same time their right or left character.

  In other words: if the ingoing string has a right (or left)
 excited mode with a given polarization  B , there will be in the outgoing
  state:
 (i)     an amplitude $\A^{AB}_{n\pm}$  for a right (or left) mode
  polarized in the A-direction and
 (ii)    an amplitude  $\B^{AB}_{n\pm}$  for a left (or right) antiparticle
mode
 polarized in the A-direction.

 In field theory, mixing of particle and antiparticle modes usually
 means pair particle production. This is not the case here.
 A string always describe a {\bf single} particle. Indeed, this
 particle is not of a fixed kind, but depends on the excitation state
 of the string. A scalar for the ground state, a vector particle for
 a one quantum excitation (in open strings), a tensor particle
 (graviton) for a two mode excitation (in closed strings), etc. [\gsws ].
 Here, the mixing (ii) of modes and antimodes \bogo\ implies
 inelastic processes changing the string excitation state
 due to the interaction with the geometry.
 In this process the initial particle of mass
  m  and spin  s  {\bf transmutes} into a different final particle of mass  m'
   and spin  s' .
 We also find elastic processes (i) in which initial and final states
  describe the same particle (although the momentum and the spin
 polarization may change) [\trans ].Concrete examples of Bogoliubov
 transformations of the type \bogo\ and the ensuing particle
 transmutations are reported below in secs. (4) and (5).
  It should be noticed that the
 particle transmutation phenomena appears at the zero genus level
 without introducing string loop corrections.

	 The Bogoliubov transformation \bogo\ can be written as
 $$
 \a^{A}_{n+} = \exp(\G) \,\a^{A}_{n-} \, \exp(-\G)\quad,\quad
\tilde \a^{A}_{n+} = \exp(\G) \,\tilde \a^{A}_{n-} \, \exp(-\G)
 \eqn\nnb
 $$
 with $ \G = - \G^{\dagger}$.
  The in and out vacua verify
 $$
\a^{A}_{n-} |O_{-}> = \tilde \a^{A}_{n-} |O_{-}> = 0 \quad ,
 \quad \a^{A}_{n+}|O_{+}> =\tilde \a^{A}_{n+}|O_{+}> = 0 ,
 \eqn\vas
 $$
 for all  $n \geq 1$. They are related by
 $$
      |O_{+}> \, = \, \exp( \G )\, |O_{-}>
 \eqn\trvas
 $$
 Since the coefficients of the transformation eq.\bogo\
  are not neccessarily real, $\G$ in eq.\bogo\  has the form:
 $$\eqalign{
	 \G  = \sum_{n=1}^{\infty} \, & (1/n)\sum_{A,B=0}^{D-1}
   \D^{AB}_{n} \a^{A}_{n-} \tilde\a^{B}_{n-}-(\D^{AB}_{n})^{*}
  (\a^{A}_{n-})^{\dagger} (\tilde\a^{B}_{n-})^{\dagger} + \cr
 & +\left[\C^{AB}_{n}-(\C^{AB}_{n})^{*} \right]
 \left[\a^{A}_{n-}(\a^{B}_{n-})^{\dagger}
 +\tilde\a^{A}_{n-}(\tilde\a^{B}_{n-})^{\dagger}\right]  \cr }
 \eqn\gexpl
 $$
	 At the first order in  $g^2$ , the coefficients $\D^{AB}_{n}$
 and $\C^{AB}_{n}$ are related with the $\A^{AB}_{n-} $ and
 $\B^{AB}_{n-} $ through
 $$\eqalign{
	 \A^{AB}_{n-} = & \, \delta^{AB} +\C^{AB}_{n}-(\C^{AB}_{n})^{*} \cr
	 \B^{AB}_{n-} = & \, (\D^{AB}_{n} )^{*}  \cr }
 \eqn\bope
 $$
 Therefore, from eqs.\trvas\ - \bope\ , the in-out vacuum transition amplitude
  is given at first order by
 $$
	 <O_{+}|O_{-}> \, = \,<O_{-}| 1 - \G |O_{-}>\,  =  1 - 4 i
 \sum_{A=0}^{D-1}\sum_{n=1}^{\infty} \Im\, \C^{AB}_{n}
 \eqn\vava
 $$
 We see that the out-vacuum follows from the in-vacuum by filling it with
  ingoing mode pairs (here a mode pair is formed by a right and a left mode).
 Conversely, the in vacuum as seen by out observers is a superposition
 of all kind of out particle states. In particular, eq.\vava\ gives the
 amplitude to find the lightest scalar particle in the out state,
 when the ingoing state was precisely this lightest scalar.
 This effect is also a manifestation of the composite
 character of the strings. The infinite set of oscillator modes,
 constituting the string, become excited during the scattering by the
  influence of the gravitational field. In fact, any localized external
 field (gravitational or not) would lead to qualitatively similar
 effects.

	 The symmetric 	and traceless part of $\B^{AB}_{1-}$  gives
 the transition amplitude from the ground state $|O_{-}> $
 to an outgoing graviton. The transition amplitude from
 $|O_{-}>$  to a final massless scalar (dilaton) is given by the trace of
   $\B^{AB}_{1-}$  and the antisymmetric part of $\B^{AB}_{1-}$
   gives the transition amplitude from
  $ |O_{-}>$  to a massless antisymmetric tensor.
 The transition from $|O_{-}>$ to a higher massive state
 $$
   \prod_{s=1}^{K} (\a^{i_{s}}_{n_{s}-})^{\dagger}
 (\tilde\a^{B}_{n_{s}-})^{\dagger} |0_{-}>
 $$
 is in general non-zero and proportional to
 $(\B^{ij}_{n-})^K$.

	 Here, we have considered bosonic strings. Particle transmutation
 takes also place for superstrings where the present discussion has been
 generalized in ref.[\trans ].
 We find that the massless particles cannot transmute
  among themselves for bosonic backgrounds (see below for fermionic
 bakgrounds). Only elastic transitions between ingoing and outgoing
  massless particles are allowed. We observe that this is a general property
 exact to all orders in $\sqrt{\a'} $, due to the fact that the transformation
 between the fermion oscillator operators  $S^{A}_{n-}$
 preserve the boson-fermion parity.
 The transitions from the massless states to a state with an
 {\bf even} number of fermion operators  $S^{A}_{n-}$   and an {\bf
 even} number of boson operators $\a^{B}_{m-}$ are non-zero.
 Transitions from the ground states
 (massless) to high massive states
 $$
   \prod_{s=1}^{K} (\a^{i}_{n_{s}-})^{\dagger}
(\tilde\a^{j}_{n_{s}-})^{\dagger}
  \prod_{r=1}^{L} (S^{A}_{n_{r}-})^{\dagger} (\tilde
S^{B}_{n_{r}-})^{\dagger}|0_{-}>
 $$
 are non-zero and proportional to  $(\B^{ij}_{n-})^K (\H^{AB}_{n-})^L$,
 where  $(\H^{AB}_{n-})$
 stand for the fermionic Bogoliubov coefficients mixing   $S^{A}_{n-}$   with
   $\tilde S^{B}_{n+}$  . It must be noticed that several features, like the
index
  structure of the Bogoliubov coefficients follow directly from
 the symmetry properties
 of the spacetime geometry. However, to find their explicit form, one must
 solve the string equations in the interaction region, close to the
 scattering center.
 In refs.[\negro ] we solved the string equations to second order in $g$
  using the expansion method (eq.\metod ) around the center of mass
 solution for the Schwarzschild black hole. We found in this way
 explicit formulas for $ \A^{AB}_{n-}$ and $\B^{AB}_{n-}$.

 In ref.[\stran ], we propose the first to introduce a background
 containing also {\bf fermionic} degrees of freedom, solution
 of the N=1 supergravity equations in D=8 dimensions. We found in
 [\stran ] an explicit  N = 1 linearized supergravity shockwave
 solution giving the spin-3/2 Rarita-Schwinger field (gravitino),
  and an exact (full non-linear) gravitational shockwave bosonic part.
  We studied then a Green-Schwartz superstring propagating in such
 supergravity background. We wrote and solved the superstring
 equations. Contrarely to the purely bosonic shockwave case (in which
 spinor-string propagation is free [\ondch]), spinor-string coordinates couple
 here non-trivially to the transverse bosonic coordinates through the
 fermionic background. We found that outgoing fermionic-modes are mixed
 with ingoing bosonic (and fermionic) modes . This leads to a
 {\bf new feature} of particle transmutation between bosons and
 fermions described by the superstring ground states (and also
 by the excited states). The presence of the bosonic-fermionic
 background provides in string theory a {\bf natural physical mechanism} to
 transform bosons into fermions (and viceversa) .

  A particular consequence of this fact is the existence of non-zero
 transitions among the superstring ground states ; this means we have
 transmutation of massless fermions into massless bosons and viceversa.
 (This massless super Yang-Mills N=1, D=10 multiplet describes physical
  particles upon supersymmetry breaking at energies much lower than the Planck
and
 grand unification scales). In ref.[\super ], we analysed and computed
  explicitly the more relevant and new particle transmutations that
 take place when an open superstring propagates through the
 supergravitational shock-wave previously found in [\stran ].Transition
 amplitudes  among the ground states, first and second excited states
 are explicitly given in ref.[\super].

 \chapter{ Exactly solvable cases of strings propagating on curved
 geometries : shock-waves and  nonlinear plane waves.}

	 We consider in this section geometries where the full
 non-linear string equations \conouno\ and constraints \conodos\ can be
 exactly solved in closed form.
 We start by considering shock-waves. They describe the geometry around
 an ultrarelativistic source propagating in a fixed direction.
 The string equations are solvable both for point like sources and for
 transversely extended sources[\ondch ]. Point-like sources correspond to an
 infinitely boosted black-hole and it is known as the Aichelburg-Sexl
 metric. The solvability of field[\camas ] and string
 equations[\ondch ] in
 such geometries follow from the everywhere flatness of the spacetime
 except on the shock-wave plane  $U \equiv  X  -  T  =  0 $ . An interesting
 non-trivial dynamics is described by the string solution which is
 obtained by matching flat spacetime string solutions across this plane.

	 The shock-wave spacetimes are described by the metric
 $$
 (ds)^2  =  - dU dV + \sum_{i=1}^{D-2} (dX^{i})^2
- \delta (U)~ f_{D}(\vec{X})~(dU)^2
 \eqn\choq
 $$
 where  U and V are null coordinates,  $U  \equiv  X^0 - X^{D-1}  ,
   V \equiv X^0 + X^{D-1}$  ,  $X^i ,\, 1 \leq i \leq D - 2 $  are
  transverse spatial coordinates and $\vec{X} = (X^{1},...,X^{D-2})$.
  The function  $f_{D}(\vec{X})$
 obeys the equation
 $$
  \partial_{i}^2  f_{D}(\vec{X})  = 16 \pi G \,\tilde p \,\, \mu (\vec{X})
 \eqn\pois
 $$
 whose solution can be written as
 $$
  f_{D}(\vec{X}) = -16 \pi G \, \tilde p \int {{d^{D-2}k} \o
 {\vec{k}^2}} {\hat \mu}(\vec{k}) \exp{ i \vec{k}. \vec{X}}
 \eqn\solpo
 $$
 where
 $$
 {\hat \mu }(\vec{k})  = \int { {d^{D-2}X} \o {(2 \pi)^{D-2}}} \, \,
  \mu(\vec{X}) \, \exp{-i \vec{k}. \vec{X}}
 $$
 and G is the gravitational constant. This metric describes the
 gravitational field of an ultrarelativistic source moving along
 the  $X^{D-1}$ axis with momentum  $\tilde p$ . $\mu(\vec{X})$
  stands for the mass density as a function of the transverse
 coordinates. For a point-like source $(\mu (\vec{X}) = \delta(\vec{X}))$
 we find the Aichelburg-Sexl
 metric (AS) :
 $$\eqalign{
 f_{D}(\rho)  = \, & K \, \rho^{4-D},\,\,  for  D > 4 \,\,  and \cr
   f_{4}(\rho)  =\,  & 8 G  \tilde p \, \log \rho ,}
 \eqn\as
 $$
 where  $\rho \equiv \sqrt{ \vec{X}.\vec{X}}$ ,
 $K \equiv -{{8 \pi^{2-D/2}}\o {D-4}} \, \Gamma(D/2 - 2) \, G \tilde p$.
 Notice that the space-time is everywhere flat except on the plane
 $U = X^0 - X^{D-1} = 0$ ,  where a shock wave is located.

	 The Riemann tensor follows from the metric \choq\ , with the
 result
 $$
 R^{UiUj}  =  (1/2) \, \delta(U)\,
 \partial_{i}\partial_{j}f_{D}(\vec{X})
 \eqn\riem
 $$
 The string equations of motion \conouno\  take the following form
 for the shock-wave \choq\
 $$\eqalign{
		 U''  -  \ddot U  = &  \, 0	\cr
 V''  -  \ddot V  +  f_{D}(\vec{X}) \left[(U')^2 - (\dot U)^2 \right]
 \delta '(U)  + &  2 \, \partial_{i} f_{D}(\vec{X}) \left[ U' X'^{i} -
 \dot U \dot X^i \right] \delta (U) =   \, 0	 \cr
 X''^{i} - \ddot X^{i} + (1/2) \, &\partial_{i}f_{D}(\vec{X})
 \left[(U')^2 - (\dot U)^2 \right] \delta(U) =  \, 0 \cr}
 \eqn\eqch
 $$
 where  $'$  and  $^{.}$  stand for ${{\partial }\o {\partial \sigma }}$
  and  ${{\partial }\o {\partial \tau }}$, respectively .

	 Since  $U(\s,\tau)$  obeys the d'Alembert equation, we can
 always make a conformal transformation $\s \pm \tau \to
 \F_{\pm}(\s \pm \tau)$ , such that
 $$
 U   =   2 \, \a' p^U  \tau
 \eqn\gauge
 $$
 where $p^U$  is constant. Therefore, the equations of motion \eqch\ become
 $$\eqalign{
 V''  -  \ddot V  -  f_{D}(\vec{X}(\s ,0))~ \delta '(\tau)  +
   \dot f_{D}(\vec{X}(\s, \tau))~ \delta (\tau) =  & \,0  \cr
 X''^{i} - \ddot X^{i} - \a' p^U \partial_{i}f_{D}(\vec{X}(\s,0))}
 \eqn\ecuac
 $$
	 The string coordinates satisfy the d'Alembert equation for all
 $(\s,\tau)$ except at  $\tau = 0$, where  $\dot X^{i}(\s,\tau)$ ,
  $V(\s,\tau)$  and  $\dot V(\s,\tau)$  are discontinuous.
 One finds from eq. \ecuac\ [\ondch ] :
 $$\eqalign{
	  \dot X^{i}_{>}(\s,0)  - \dot X^{i}_{<}(\s,0)   =&
   -  \a' p^U \, \partial_{i}f_{D}(\vec{X}(\s,0)) \cr
		  X^{i}_{>}(\s,0)  -  X^{i}_{<}(\s,0)    =&   \, 0 \cr}
 \eqn\empal
 $$
 Here $ X^{i}_{<}(\s,\tau) $ and  $X^{i}_{>}(\s,\tau)  $  stand for the string
 coordinates before  ($\tau < 0$) and after ($\tau > 0$) the collision
 with the shock wave. Notice that  $U_>  =  U_<  =  2  \a' p^U  \tau$
 is continuous at  $\tau = 0$. The last term in the first equation in
 \ecuac\  is indeed ambiguous due to the discontinuity of
 $\dot X^{i}(\s,\tau)$  at  $\tau = 0$.
 This ambiguity has been solved in ref.[\costa ] .

 Since  $X^{i}_{<}(\s,\tau)$  and  $X^{i}_{>}(\s,\tau)$ obey
 the flat space-time
 (d'Alembert) equations, we can write the string solutions as in
 Minkowski spacetime [cfr. eq.\solm ]
 $$\eqalign{
 &X^{A}_{<}(\s,\tau)  =  q^{A}_{<} \, +\, 2 p^{A}_{<} \a' \tau ~ + \cr
 &+i \sqrt{\a'} \sum_{n \neq 0}
   \{ \a^{A}_{n<} \exp[in(\s - \tau)]
	 + \tilde  \a^{A}_{n<} \exp[-in(\s + \tau )] \}/n
 \quad for \, \tau < 0 \cr
 &X^{A}_{>}(\s,\tau)  =  q^{A}_{>} \, + 2 \, p^{A}_{>} \a' \tau ~ + \cr
 &+i \sqrt{\a'} \sum_{n \neq 0}
   \{ \a^{A}_{n>} \exp[in(\s - \tau)]
	 + \tilde  \a^{A}_{n>} \exp[-in(\s + \tau )] \}/n
 \quad for \, \tau > 0 \cr}
 \eqn\smen
 $$
 Inserting eqs.\smen\  in the discontinuity relations \empal\
  yields the following matching relations:
 $$\eqalign{
	 q^{i}_{<} -q^{i}_{>} =   ~ 0 ~  ,	~
 p^{i}_{>}  - p^{i}_{<}   =& \, {{p^U \tilde
K}\o{2\pi}}\int_{0}^{2\pi} d\s  D^{i}(\s) \, ,\,
 q^{V}_{>}  -   q^{V}{<}   =	- {{K}\o{2\pi}}\int_{0}^{2\pi}  d\s
 C^{4-D}(\s) \,
 \cr
 \a^{A}_{n>} - \a^{A}_{n<} =& \,   {{\sqrt{\a'} p^U \tilde K}\o{2\pi}}
 \int_{0}^{2\pi} d\s  D^{i}(\s) \exp(in\s)	\, ,\cr
 \tilde\a^{A}_{n>} - \tilde\a^{A}_{n<} =& \,  {{\sqrt{\a'} p^U \tilde
K}\o{2\pi}}
 \int_{0}^{2\pi}  d\s  D^{i}(\s) \exp(-in\s) \, ,
  \cr}		\eqn\resol
 $$
 where
 $$
	 C^{i}(\s)   =  q^{i} + i \sqrt{\a'} \sum_{n \neq 0}
   \{ \a^{i}_{n<} - \tilde  \a^{A}_{-n<} \} \exp[-in\s] /n \quad ,\>
 \tilde K \equiv {{D-4} \o 2} K
 \eqn\defc
 $$
 , $C(\s) \equiv \sqrt[C^{i} C^{i}]$ and
 $$
 D^{i}(\s)  = -{1 \o {2 \tilde K}}{{\partial f_{D}(\vec{C})}\o{\partial C^{i}}}
 \eqn\defd
 $$
 The constraint equations simply read here
 $$
 \pm \partial_{\pm}V_{<}  =   {1 \o {\a' p^{U}}}\,
 (\partial_{\pm}X^{i}_{<})^2 \quad for \, \tau<0
   ~   and  ~
 \pm \partial_{\pm}V_{>}=  {1 \o {\a' p^{U}}} \,
 (\partial_{\pm}X^{i}_{>})^2 \quad for \, \tau > 0 .
 \eqn\vipo
 $$
 Hence, the coordinate  $V(\s,\tau)$  is not
 an independent variable and can be obtained from the transverse
 coordinates  $X^{i}(\s,\tau)$  through
 eqs.\vipo\ up to the center of mass position $q^V$ . $q^V$ is an
 independent dynamical variable, canonically conjugate to $p^U$ (see
 [\costa ] for details).

	 We give a well defined meaning to the transformation \resol\
  at the quantum level by using the integral representation [\plb
 ,\costa ,\eneimp ]
 $$
 D^{i}(\s) = -{{2i \pi^{D/2-1}} \o { \Gamma(D/2-1)}} \int {{d^{D-2}k}
 \o {\vec{k}^2}} \tilde \mu(\vec{k}) :\exp{i\vec{k}.\vec{C}(\s)}: \, k^i
 \eqn\inre
 $$
 where  $: ... :$  stands for normal ordering with respect to the
 operators  $ \a^{i}_{n<}$ , $\tilde\a^{A}_{n<}$. Using this formula,
 the transformation from operators  $<$  to operators  $>$  can be
 generated by  $\exp(\G)$  where
 $$
 \G   =  -2 i G p^U \int_{0}^{2\pi} d\s \int {{d^{D-2}k}
 \o {\vec{k}^2}} : \exp{i\vec{k}.\vec{C}}: \tilde \mu(\vec{k})
 \eqn\gech
 $$
 As shown in ref.[\eneimp ] , $\exp(\G)$  is unitary provided the
 integrals $\int d^{D-2}k$ are taken in principal value.
	 In this framework, we have been able to compute physical
 quantities like the string squared mass [\plb ,\eneimp ],
 $$
 \a'(M_{>})^2 = -{{D-12} \o {12}} + \sum_{n=1}^{\infty}\sum_{i=1}^{D-2}
 \left[(\a^{i}_{n>})^{\dagger}\a^{i}_{n>} + (\tilde\a^{i}_{n>})^{\dagger}
 \tilde\a^{i}_{n>}\right],
 \eqn\mas
 $$
 the number of modes N and the energy-momentum tensor [\eneimp ].
 All these quantities turn to be {\bf finite} in shock-wave space-times.
 That is, there is no need of renormalization in contrast with the
 case in quantum field theory.
	 The expectation value of the string mass  $(M_{>})^2$
 after the collision with the shock-wave happens to be calculable
 in closed form. We found for the string ground-state  $|O_{<}>$
   (the tachyon) [\plb ,\eneimp ]:
 $$
  {{< 0_{<}| (M_{>})^2 |0_{<} >} \o {<0_{<}|0_{<} >}} = \mu_{0}^2
   -  32 \mu_{0}^2 L^{2-D} G^2 \pi^{3/2} \int d^{D-2}k
 |\mu(\vec{k})|^2 {{\tan{[\pi \a'\vec{k}^2]}  \Gamma(\a'\vec{k}^2) \o
  \vec{k}^2 \, 2^{2\a'\vec{k}^2} \Gamma(1/2 + \a'\vec{k}^2) }}
 \eqn\masc
 $$
 where  $\mu_{0}^2 = -{{D-12} \o {12 \a'}}$  is the tachyon squared
 mass. This term is the initial mass of the string. The second term
 in eq.\masc\ describes the the change of the string squared mass
 after the interaction with the shockwave. It can be interpreted
 as the infinite superposition of particle states that form
 $|0_{-}>$ as seen by the outgoing (+) observers. Notice that this
 term is proportional to $L^{D-2}$. That is the transverse size
 of the shockwave.
 Analogous formulae can be derived for excited states.

 It should be noticed that the integrand in eq.\masc\  possess
 real singularities which we integrate
 in principal value following the prescription that makes $\exp(\G)$
 unitary. These equally spaced real poles in  $\vec{k}^2$  are
 characteristic of the tree level string spectrum (real mass
 resonances)[\gsws ]. The presence of such poles is not related at all
 to the structure of the spacetime geometry (which may be or not
 singular). It should be noticed that the claim in ref.[\horo ] that
 $<0_{<}|(M_{>})^2|0_{<}>$  is infinite in shock-wave spacetimes was incorrect
 as shown in refs.[\plb ]. The only divergences that may eventually
 appear are related to the infinite transverse size of the shockwaves
 , as the factor $L^{D-2}$ in eq.\masc\ .

	 The exact expectation value of all the components of the
 energy-momentum tensor in shock-wave spacetimes of the type \choq\
 were computed in ref.[\eneimp ]. We expressed all of them in terms
 of explicit integral representations. All these physical magnitudes
 turned to be {\bf finite} except at $\tau = 0$. That is, they are
 finite except if one sits exactly at the singularity.
 The finiteness of these string calculations should be contrasted
 with QFT where the energy-momentum tensor always need
 regularization and renormalization [\bida ].

	 In ref.[\costa ] the two point amplitude,  $A_{2}(
 \vec{k}_{2},\vec{k}_{1}) $ ,
 describing the scattering of the lowest string excitation (the
 tachyon) by a gravitational shock wave of the type \choq\  is computed.
 For this purpose, we used the appropiate vertex operator in this
 background. We explicitly evaluated in ref.[\costa ],
  $A_{2}(\vec{k}_{2},\vec{k}_{1}) $   for
 large impact parameters  q. It is given by the Coulombian amplitude
 plus string corrections of order  s/q  [ $s = (k^1 + k^2)^2$ ], for
 large q. These string corrections produce an infinite sequence of
 imaginary poles in s, the semi-infinite sequence of Coulomb poles
 noticed by 't Hooft [\gth ] remaining always present.

 Let us now consider strings propagating in
 gravitational plane-wave space-times. These are sourceless
 gravitational fields described by the metric
 $$
 (ds)^2  =  - dU dV + \sum_{i=1}^{D-2}(dX^{i})^2  - \left[ W_{1}(U)\, (X^2 -
Y^2 )
 + 2 \,W_{2}(U) \,X Y \right]\, (dU)^2
 \eqn\opla
 $$
 where $ X \equiv X^1$ , $Y \equiv X^2$ .  These space-times are exact
 solutions of the vacuum Einstein equations for any choice of the
 profile functions $ W_{1}(U)$ and $W_{2}(U)$.
  In addition they are exact string vacua [\guven ].
  The case when $W_{2}(U) = 0$  describes waves of constant
 polarization. When both $ W_{1}(U) \neq 0$ and  $W_{2}(U)\neq 0$ ,
 eq.\opla\ describes waves with arbitrary  polarization.
  If $W_{1}(U)$ and/or  $W_{2}(U)$ are singular functions, space-time
 singularities will be present. The singularities will be located on
 the null plane  U = constant. We consider profiles which are nonzero
 only on a finite interval  $-T < U < T$ , and which have power-type
  singularities [\ondpl ],
 $$
 W_{1}(U) = { {\a_{1}} \o {|U|^{\beta_{1}}}} \quad , \quad
 W_{2}(U) = { {\a_{2}} \o {|U|^{\beta_{2}}}}
 \eqn\singu
 $$
 The spacetimes \opla\ share many properties with the shockwaves
 \choq\ . In particular, $U(\s,\tau)$ obeys the d'Alembert equation
 and we can choose the light-cone gauge \gauge\ . The string equations
 of motion \conouno\ become then in the metric \opla\ :
 $$\eqalign{
 V''  -  \ddot V ~ + ~&(2 \a' p^U )^2 \left[ \partial_{U} W_{1}\,(X^2 - Y^2 )
 ~ +~2 \,\partial_{U}W_{2} \,X Y \right]\cr
  + &~8 \a' p^U \left[ W_{1} ( X \dot X - Y \dot Y ) +W_{2} (X \dot Y + \dot Y
X )
 \right] =  \,0  \cr
 X'' - \ddot X ~ +~ &(2 \a' p^U )^2 \left[ W_{1} X -W_{2} Y \right] = \,0
 \cr Y'' - \ddot Y ~ +~ &(2 \a' p^U )^2 \left[ W_{2} X -W_{1} Y \right] = \, 0
\cr}
 \eqn\equac
 $$
 and the constraints \conodos\ take the form:
 $$
 \pm \partial_{\pm}V_{<}  =   {1 \o {\a' p^{U}}}
 \left\{ (\partial_{\pm}X)^2 +
 (\partial_{\pm}Y)^2 + \sum_{i=3}^{D-2}(\partial_{\pm}X^{i})^2
  \right\} + \a' p^{U} \left[  W_{1}\, (X^2 - Y^2 )
 + 2 \,W_{2} \,X Y \right]
 \eqn\vinop
 $$
 Let us analyse now the solutions of the string equations \equac\ and \vinop\
 for a closed string. The transverse coordinates obey the d'Alembert equation
 , with the solution
 $$
	 X^{i}(\s,\tau)  =  q^i + 2 p^i \a' \tau + i \sqrt{\a'} \sum_{n \neq 0}
   \{ \a^{i}_{n} \exp[in(\s - \tau)]
	 + \tilde  \a^{i}_{n} \exp[-in(\s + \tau )] \}/n ,~ 3 \leq i \leq D-2
 \eqn\stra
 $$
 For the X and Y components it is convenient to Fourier expand as
 $$
 X(\s,\tau) =  \sum_{n=-\infty}^{+\infty} \exp(in\s) ~ X_{n}(\tau)\quad , \quad
 Y(\s,\tau) =  \sum_{n=-\infty}^{+\infty} \exp(in\s) ~ Y_{n}(\tau)
 \eqn\expf
 $$
 Then, eqs.\equac\ for X and Y yield
 $$\eqalign{
  \ddot X_{n} + n^2 X_{n}-&(2 \a' p^U )^2
 \left[ W_{1} X_{n} - W_{2} Y_{n} \right]= ~ 0 \cr
  \ddot Y_{n} + n^2 Y_{n}-&(2 \a' p^U )^{2}
 \left[ W_{2} X_{n} - W_{1} Y_{n}\right]= ~ 0 \cr}
 \eqn\schr
 $$
 where we consistently set $ U  = 2 \a' p^U \tau$.
 Formally, these are two coupled one-dimensional Schr\"odinger-like
 equations with $\tau$ playing the r\^ole of a spatial coordinate.

 We study now the interaction of the string with the gravitational
 wave. For  $2\a' p^{U} \tau < -T$, $W_{1,2}(\tau) = 0$ and therefore  X, Y are
 given by the usual flat-space expansions
 $$\eqalign{
 X(\s,\tau)  =~ & q^X_{<} + 2 p^X_{<} \a' \tau + i \sqrt{\a'} \sum_{n \neq 0}
   \{ \a^{X}_{n<} \exp[-in\tau)]
	 - \tilde  \a^{X}_{-n<} \exp[in\tau )] \} \exp[in\s]/n \cr
 Y(\s,\tau)  =~ & q^Y_{<} + 2 p^Y_{<} \a' \tau + i \sqrt{\a'} \sum_{n \neq 0}
   \{ \a^{Y}_{n<} \exp[-in\tau)]
	 - \tilde  \a^{Y}_{-n<} \exp[in\tau )] \} \exp[in\s]/n \cr}
 \eqn\inso
 $$
 These solutions define the initial conditions for the string
 propagation in $\tau \geq -\tau_{0} \equiv -{T \o {2\a' p^{U}}}$.
 In the language of the Schr\"odinger-like equations we have a
 two channel potential in the interval $-\tau_{0} < \tau < +\tau_{0}$.
 We consider the propagation of the string when it approaches the
 singularity at $U = 0 = \tau$ from $\tau < 0$. When $ W_{1}$ is more singular
 at $U = 0$ than $W_{2}$, i.e. $\beta_{1} > \beta_{2}$ in eq.\singu ,
 the string behaviour is determined by $W_{1}$. Let us consider the
 case when both singularities
 are of the same type; i.e. $\beta_{1} = \beta_{2} \equiv \beta $.
 This case is actually generic
 since the case when  $W_{2} ( W_{1} )$  is more singular than
 $W_{1}( W_{2})$ can be
 obtained by setting $\a_{1} = 0 ~(\a_{2} = 0)$ in the $\beta_{1} =
 \beta_{2}$ solution.

 Eq.\schr\ can be approximated near $\tau = 0^-$
  as
 $$\eqalign{
 \ddot X_{n}-{{(2 \a' p^U )^{2-\beta}}\o{|\tau|^{\beta}}}
 \left[ \a_{1} X_{n} +\a_{2} Y_{n}\right]=&0 \cr
 \ddot Y_{n}-{{(2 \a' p^U )^{2-\beta}}\o{|\tau|^{\beta}}}
 \left[ \a_{2} X_{n} -\a_{1} Y_{n}\right]=&0 \cr}
 \eqn\aprox
 $$
 The behaviour of the solutions $X_{n}(\tau)$ and $Y_{n}(\tau)$
 for $\tau \to 0$ depends crucially on the value range of $\beta$. Namely,
 i) $\beta > 2$,
 ii) $\b = 2$,
 iii) $\beta < 2$.
 For simplicity, we start with the case $\beta = 2$ where the solution
 is [\ondpl ]:
 $$
 X_{n}(\tau)\buildrel{\tau\to 0^{-}}\over = C |\tau|^{\lambda}\quad ,
 \quad Y_{n}(\tau)\buildrel{\tau\to 0^{-}}\over = D |\tau|^{\lambda}
 \eqn\tauce
 $$
 Which has four solutions:
 $$
 \lambda_{1,2} = {1\o 2}[1 \pm \sqrt{1 - 4 \tilde \a}]\quad , \quad
 \lambda_{3,4} = {1\o 2}[1 \pm \sqrt{1 + 4 \tilde \a}]
 \eqn\lamda
 $$
 Here, $\tilde \a \equiv \sqrt{ \a_{1}^{2} + \a_{2}^{2}}$.
 Notice that for any real value $\a_{1}$ and $\a_{2}$ , we have
 $\lambda_{4} < 0$ , $\lambda_{3} > 0$ , $Re \lambda_{1,2} > 0 $.
 The solution associated with $\lambda_{4}$ diverges at $\tau = 0$ as
 $$
 X_{n}(\tau)_{4}\buildrel{\tau\to 0^{-}}\over = c~
 |\tau|^{\lambda_{4}}~(\a_{1} + \tilde \a ) \quad ,
 \quad Y_{n}(\tau)_{4}\buildrel{\tau\to 0^{-}}\over = c ~\a_{2}~
 |\tau|^{\lambda_{4}}
 \eqn\tce
 $$
 where c is an arbitrary constant.
 The solutions associated with $\lambda_{1,2}$ and $\lambda_{3}$
 vanish for $\tau \to 0^{-}$.
 $$\eqalign{
 X_{n}(\tau)_{3}\buildrel{\tau\to 0^{-}}\over = d~
 |\tau|^{\lambda_{3}}~(\a_{1} + \tilde \a ) \quad &,
 \quad Y_{n}(\tau)_{3}\buildrel{\tau\to 0^{-}}\over = d ~\a_{2}~
  |\tau|^{\lambda_{3}}
 \cr X_{n}(\tau)_{1,2}\buildrel{\tau\to 0^{-}}\over = k_{1,2}~
 |\tau|^{\lambda_{1,2}}~(\a_{1} - \tilde \a ) \quad &,
 \quad Y_{n}(\tau)_{1,2}\buildrel{\tau\to 0^{-}}\over = k_{1,2}~ \a_{3}
  ~|\tau|^{\lambda}_{1,2} \cr}
 \eqn\tosci
 $$
 Here d, $ k_{1}$ and $ k_{2}$ are arbitrary constants.
 For  $\tilde \a > 1/4$, the solutions $ X_{n}(\tau)_{1,2}$ and
 $Y_{n}(\tau)_{1,2}$  approach the singularity oscillating with
 decreasing amplitudes.
 As it is clear, for generic initial conditions, the string behaviour
 near $\tau \to 0^- $ is dominated by the $(X_{n}(\tau)_{4} ,
 Y_{n}(\tau)_{4})$ solution.
 The fact that $(X_{n}(\tau)_{4} , Y_{n}(\tau)_{4})$ diverges when
 $\tau \to 0^- $, means that
 the string goes to infinity as it approaches the singularity plane.
 From eqs.\tce\ , we see that the string goes to infinity in a direction
  forming an angle $\a$ with the X-axis in the X, Y plane. The string
 escape angle $\a$ is given by
 $$
 \tan \a = { {\a_{2}} \o {\tilde \a +  \a_{1}}} \quad i. e.
 \quad \tan 2\a = {{ \a_{2}} \o { \a_{1}}}
 \eqn\esca
 $$
 (see fig.1).
 When $ \a_{2} = 0$ , then $\a = 0$ and the string escapes to infinity
 in the X-direction. For $ \a_{1} = 0$ , then $\tan \a = sign( \a_{2}) =
 \pm 1$ and the string goes to infinity with an angle $(\pi/4) sign (\a_{2})$.
  For $ \a_{1} > 0$,  and arbitrary   $\a_{2}$, the allowed directions
 are within the cone  $|\a|<\pi/4$ , as depicted in fig.1.
 If $ \a_{1} < 0$ the string escape angle is
 within the cone $|\a-\pi/2| < \pi/4$. In summary, for arbitrary values of
 $ \a_{1}, \a_{2}$ , the string escape angle can take any value between
 0 and $2 \pi$.

	 Let us now consider the case $\beta > 2$. We have [\ondpl ]
 $$
 X_{n}(\tau)\buildrel{\tau\to 0^{-}}\over = C [\a_{1} \pm \tilde \a]~
 |\tau|^{{\beta}\o 4} ~ \exp[K|\tau|^{1-\beta/2}]    \quad ,
 \quad Y_{n}(\tau)\buildrel{\tau\to 0^{-}}\over = C \a_{2} ~
 |\tau|^{{\beta}\o 4} ~ \exp[K|\tau|^{1-\beta/2}]
 \eqn\bema
 $$
 where C is an arbitrary constant K can take the four following values:
 $$
 K_{1,2} = \mp i\,{{(2 \a' p^{U} )^{1-\beta/2}}\o
 {\beta/2 - 1}}\sqrt{\tilde \a} \quad , \quad
 K_{3,4} = \mp {{(2 \a' p^{U} )^{1-\beta/2}}\o
 {\beta/2 - 1}}\sqrt{\tilde \a}
 \eqn\kaes
 $$
 As for the $\beta = 2$ case, for $\beta \geq 2$ we have a divergent solution
for
 $\tau \to 0^- $ associated to $K_4 $:
 $$\eqalign{
 X_{n}(\tau)_{4}\buildrel{\tau\to 0^{-}}\over = &C \, [\a_{1} + \tilde \a]
 \, |\tau|^{{\beta}\o 4} \exp\left[{{(2 \a' p^{U}|\tau| )^{1-\beta/2}}\o
 {\beta/2 - 1}}\sqrt{\tilde \a}\right]    ~,~ \cr
  Y_{n}(\tau)_{4}\buildrel{\tau\to 0^{-}}\over = &C \, \a_{2}
 \, |\tau|^{{\beta}\o 4} \exp\left[{{(2 \a' p^{U} |\tau|)^{1-\beta/2}}\o
 {\beta/2 - 1}}\sqrt{\tilde \a}\right] \cr}
 \eqn\pun
 $$
 The other three solutions vanish
 for $\tau \to 0^- $. The solutions (1,2) oscillate for $\tau \to 0^- $
 with decreasing amplitude.
	As in the $\beta = 2$ case, and for generic initial conditions,
 the string behaviour for $\tau \to 0^- $  is dominated by the
 $(X_{n}(\tau)_{4} , Y_{n}(\tau)_{4})$
 solution. The string goes to infinity in the same way as for the
 $\beta = 2$ case and with the same escape angle given by eq.\esca .

	 Finally, let us discuss now the situation when $\beta < 2$. In this case,
 the solution for $\tau\to 0^{-}$ behaves as
 $$
 X_{n}(\tau)\buildrel{\tau\to 0^{-}}\over = A_{1} + A_{2}~ \tau +
O(|\tau|^{2-\beta})\quad , \quad
 Y_{n}(\tau)\buildrel{\tau\to 0^{-}}\over = B_{1} + B_{2}~ \tau +
O(|\tau|^{2-\beta})~~,~~\beta \neq 1
 \eqn\beme
 $$
 [In the special case $\beta = 1$ one should add a term  $0 ( \tau
\ln|\tau|)$].
 For  $\beta < 2$, the string coordinates X, Y are always regular
 indicating that the string propagates smoothly through the
 gravitational-wave singularity U = 0. (Nevertheless, the
 velocities $\dot X$ and $\dot Y$ diverge at $\tau = 0$ when $1 \leq \beta <
2$).

	 Let us now summarize the string behaviour near
 the singularity $\tau \to 0^- $ for $\beta \geq 2$.
	 For generic initial conditions, we
 see  from eqs.\tce\ and \pun\ that the string behaves as
 $$\eqalign{
 (X(\s,\tau),Y(\s,\tau))_{\beta=2}\buildrel{\tau\to 0^{-}}\over =&
 {\cal A }(\s) ~ \sqrt{|\tau|}^
 {1-\sqrt{1+4\tilde\a}} ~~ (\a_{1} + \tilde \a,\a_{2}) ~ , \cr
 (X(\s,\tau),Y(\s,\tau))_{\beta > 2}\buildrel{\tau\to 0^{-}}\over =&
 {\cal B }(\s) ~ |\tau|^{\beta/4} ~ \exp\left[{{(2 \a' p^{U}|\tau|
 )^{1-\beta/2}}\o{\beta/2 - 1}}\sqrt{\tilde \a}\right]
 ~~(\a_{1} + \tilde \a,\a_{2})\cr}
 \eqn\bestia
 $$
 The functions ${\cal A}(\s), {\cal B}(\s)$ depend on the initial conditions.
 The above solutions imply that the string {\bf does not cross} the U = 0
 singularity plane. The string goes off to infinity in the (X,Y) plane,
 grazing the singularity plane U = 0 (therefore never crossing it).
 Here, the string escapes to infinity with an angle $\a$ given by
 eq.\esca\ that depends upon
 the strenghts of the profile singularities,  $\a_{1}$ and $\a_{2}$,
 but not on $\beta$. (This means that the escape angle is solely
 determined by the polarization of the nonlinear gravitational wave
 \opla ).
 At the same time, the presence of the oscillatory modes \tosci\ and
 \bema\ with $K_{1,2}$ imply that the string oscillates in the XY plane
 perpendicularly to the escape direction, with an amplitude vanishing
 for $\tau \to 0^- $. The non-oscillatory modes in \tosci\ and  \bema\ with
 $K_{3}$ are
 negligible since they are in the same direction as the divergent
 solutions \tce\ and \pun .

	 The spatial string coordinates $X^{i}(\s,\tau)~ [3 \leq i \leq D-2]$
 behave freely [eq.\stra ].
	 The longitudinal coordinate $V(\s,\tau)$ follows from the
 constraint eqs.\vinop\ and the solutions \bestia\  for $X(\s,\tau),
 Y(\s,\tau)$ and  $X^{j}(\s,\tau)~ [3 \leq j \leq D-2]$. We see that
 for $\tau \to 0^- $ , $V(\s,\tau)$
 diverges as the square of the singular solutions \bestia\ .

   Let us consider the spatial length element of the string, i.e. the
 lenght at fixed  $U = 2 \a'p^{U} \tau $  , between  two points
 $(\s,\tau)$  and  $(\s+d\s,\tau)$,
 $$
 ds^2 = dX^2 + dY^2 + \sum_{j=3}^{D-2} (dX^{j})^2
 \eqn\inter
 $$
 For $\tau\to 0^{-}$ eqs.\bestia\  yield
 $$\eqalign{
 ds^2\buildrel{\tau\to 0^{-}}\over =&[(\a_{1} + \tilde \a )^2 + \a_{2}^{2}]
 ~ {\cal B' }(\s)^{2} d\s^2 ~~ |\tau|^{1-\sqrt{1+4\tilde\a}}\quad for ~
 \beta=2 \, , \cr
 ds^2\buildrel{\tau\to 0^{-}}\over =&[(\a_{1} + \tilde \a )^2 + \a_{2}^{2}]
 ~ {\cal B' }(\s)^{2} d\s^2 ~~ |\tau|^{\beta/2} ~ \exp\left[{{(4 \a'
p^{U}|\tau|
 )^{1-\beta/2}}\o{\beta/2 - 1}}\sqrt{\tilde \a}\right]
 \quad for ~ \beta \geq 2. \cr}
 \eqn\explo
 $$
 That is, the proper length between $(\s_{0},\tau)$ and $(\s_{1},\tau)$ is
given by
 $$\eqalign{
 \Delta s \buildrel{\tau\to 0^{-}}\over =&
 \sqrt{(\a_{1} + \tilde \a )^2 + \a_{2}^{2}} ~
 [{\cal B }(\s_{1}) - ~{\cal B }(\s_{2})]~ \sqrt{|\tau|}^
 {1-\sqrt{1+4\tilde\a}}\quad for ~ \beta=2 \, , \cr
 ~\Delta s \buildrel{\tau\to 0^{-}}\over =&
 \sqrt{(\a_{1} + \tilde \a )^2 + \a_{2}^{2}} ~
 [{\cal B }(\s_{1}) - ~{\cal B }(\s_{2})]~~ |\tau|^{\beta/4} ~
 \exp\left[{{(2 \a' p^{U}|\tau|)^{1-\beta/2}}\o{\beta/2 - 1}}
\sqrt{\tilde \a}\right]
 ~ for ~ \beta \geq 2. \cr}
 \eqn\strech
 $$
 We see that  $\Delta s \to \infty$ for $\tau\to 0^{-}$ . That is,
 the string stretches infinitely when it approaches the singularity plane.
 This stretching of the string proper size also occurs for $\tau \to 0$
 in the inflationary cosmological backgrounds (see sec.6).

	 Another consequence of eqs.\bestia\  is that the
 string reaches infinity in a finite time $\tau$. In particular, for
 $\s$-independent coefficients, eqs.\bestia\  describe geodesic
 trajectories. The fact that for  $\beta \geq 2$, a point particle
 (as well as a string) goes off to infinity in a finite $\tau$ indicates
 that the space-time is singular.

	 What we have described is the $\tau\to 0^{-}$ behaviour for
 generic initial data. In particular, there is a class of solutions
 where  ${\cal A}(\s) \equiv 0$  and  ${\cal B}(\s) \equiv 0$ , whereas
 the coefficients of the regular modes are arbitrary.
 In this class of solutions,  X  and
 Y  vanish for  $\tau\to 0^{-}$ . However, when continued to $\tau > 0$, these
 solutions are complex. The real valued physical solution is  X = Y = 0
 for all $\tau \geq 0$. It means that the string gets trapped in the
 singularity plane  U = 0  at the point X = Y = 0 , where the
 gravitational forces are zero.

  Finally, we would like to remark that the string evolution near the
space-time
 singularity is a {\bf collective motion} governed by the nature of the
 gravitational field. The (initial) state of the string fixes the overall
 $\s$-dependent coefficients
 ${\cal A}(\s),{\cal B}(\s)$ [see eqs.\bestia  ], whereas the
 $\tau$-dependence is fully determined by the space-time geometry. In other
 words, the $\tau$-dependence is the same for all modes $n$ . In some
 directions, the string collective propagation turns to be an
 infinite motion (the escape direction), whereas in other directions,
 the motion is oscillatory, but with a fixed (n-independent) frequency.
  In fact, these features are not restricted to singular gravitational
  waves, but {\bf are generic} to strings in strong gravitational
 fields [see sec.(6) and refs.(\ondch ,\desi ,\vene )].

 For sufficiently weak spacetime singularities
  ($\beta_{1} < 2$ and $\beta_{2} < 2$), the string crosses the
 singularity and reaches the region $U > 0$. Therefore, outgoing
 scattering states and outgoing operators can be defined in the region
 $U > 0$. We explicitly found in ref.[\ondpl ] the transformation
 relating the ingoing and
 outgoing string mode operators. For the particles described by the
 quantum string states, this relation implies two types of effects
 as described in sec.3 for generic asymptotically flat spacetimes: (i)
  rotation of spin polarization in the (X,Y) plane, and (ii)
 transmutation between different particles. We computed in [\ondpl ]
  the expectation values of the outgoing mass  $M_{>}^{2}$  operator
 and of the mode-number operator $N_{>}$, in the ingoing ground state
$|O_{<}>$.
 As for shockwaves (cfr. eq.\masc ) , $M_{>}^{2}$  and $N_{>}$ have
 different expectation values  than  $M_{<}^{2}$  and  $N_{<}$ .
  This difference is due to the excitation of the string modes after
 crossing the space-time singularity. In other words, the string state
 is not an eigenstate of $M_{>}^{2}$, but an infinity superposition
 of one-particle states with different masses. This is a consequence
 of the particle transmutation
 which allows particle masses different from the initial one ($\mu_{o}^{2}$).
\chapter{ Strings in Conical Spacetimes (the geometry around a cosmic
 string)}

 Conical spacetimes describe the
 geometry around a cosmic string. The spacetime surrounding such object
 is locally flat and there is a defect angle $\a$  proportional to the
 mass density of the cosmic string. Therefore, the interaction of
 strings (or particles) with such a geometry comes from the unusual
 periodicity requirement by an angle  $2\pi  -  \a$  around the cosmic
 string [\conico ].

 A conical space-time in D-dimensions is defined by the metric
 $$
 (ds)^2  = - (dX^{0})^2  + (dR)^2 + R^2 ~(d\Phi )^2 +
 \sum_{i=3}^{D-1}(dX^{i})^2
 \eqn\conic
 $$
 where $R =  \sqrt{ X^2  +  Y^2}$ and $\Phi =  \arctan( Y / X )$
 are cilindrical coordinates, but with the range
 $$
 0 \leq \Phi < 2\pi\a  \quad ,\quad   \a  \equiv  1  -  4 G~\mu
 \eqn\defec
 $$
 $\mu$ is the cosmic string tension ($G \mu \approx  10^{-6}$
  for the Grand Unified theories cosmic strings), $(dX^{i})^2$  is a
 flat  (D - 3)  dimensional space and  $X^i , 3 \leq  i \leq D - 1$ ,
  are cartesian coordinates. The spatial points $(R, \Phi , X^{i})$
   and  $(R, \Phi + 2\pi\a,X^{i})$   are identified. The space-time
 is locally flat for  $R \neq  0$  but has a
 conelike singularity at  R = 0  with azimuthal deficit angle
 $$
 2\, \Delta  =  2  \pi \, ( 1  -  \a )   =   8 \pi\, G \mu
 \eqn\defi
 $$
    This geometry describes a straight cosmic string of zero thickness.
 It is a good approximation for very thin cosmic strings with large
 curvature radius. Globally, it has a non-trivial (multiply-connected)
topology.
	 The string equations of motion \conouno\ are free equations in the
 coordinates $X^{0}, X, Y, X^{i}$  but with the condition that the points
 $(R, \Phi, X^{i})$  and $(R, \Phi + 2\pi\a,X^{i})$  are identified. Therefore,
we
  can choose the light-cone gauge
 $$
  U  =  2\, \a'\, p^U \tau \quad , ~{\rm where}  \quad U \equiv X^0  - X^{D-1}
 \eqn\gauco
 $$
 The constraints \conodos\ in the light-cone gauge \gauco\ completely
 determine the longitudinal $V \equiv X^0 + X^{D-1}$  coordinate of the
 string in terms of the transverse coordinates, as it should be.

	 Before solving the string equations, let us consider the point
 particle propagation. Since the equations of motion are locally those
 of Minkowski spacetime, the trajectories are straight lines. The
 non-trivial point is to impose the  $2 \pi \a$  periodicity condition on the
 angle $\Phi$ . The solution is depicted in fig. 2 in coordinates where
 half of the defect angle  ( $\Delta$ )  is taken to the right and half to the
 left of the conical singularity at O. Therefore, particles passing on
 the right (left) of  O  are deflected by $+\Delta (-\Delta)$. Notice that this
 deflection angle is independent from the impact parameter as well as
 from the energy of the ingoing particle. This shows the purely
 topological nature of this infinite range interaction. The solution
 of the string equations of motion follows analogously [\conico ].
	 There appear essentially two different situations for the
 interaction between the fundamental string and the conical spacetime:

	 (i) The string does not touch the scatterer body (here
 represented by the cosmic string); in this case the string only
 suffers a deflection at the origin. We refer to this situation as
 elastic scattering.

	 (ii) The string collides against the scattering center.
 In this case, the string gets its normal modes excited besides being
 deflected. This happens each time a point of the string collides with
 the center. We refer to this process as inelastic scattering.

	 Let us first consider the case (i)  (elastic scattering). For
 $\tau \to -\infty$, the ingoing-solution $ X_{<}^{A}$ is just the free
 solution without deflection. For  $\tau \to +\infty$ , the outgoing solution
 $X_{>}^{A}$  is the free solution after deflection. We have
 $$
 (X_{>} , Y_{>} )  =  (  X_{<} , Y_{<} ) ~  {\cal R}( \pm \Delta) ~~ ,
 \eqn\rota
 $$
 ${\cal R}(\Delta)$  being the rotation matrix,
 $$
 {\cal R}(\Delta) = \pmatrix{ \cos\Delta &\sin \Delta \cr -\sin \Delta &\cos
 \Delta \cr} \quad .
 \eqn\matriz
 $$
 $\Delta$  is the deflection angle given by eq.\defi\ and the + (-)
 sign refers to a string passing to the right (left) of the cosmic
 string. The $(D-3)-X^{i}$ components are not affected by the scattering.
    From eqs.\rota\ and \matriz\ we find that the outgoing zero-modes
 and the oscillators are given by the respective in operators rotated
 by  ${\cal R}(\pm \Delta)$ .
	 In other words, there is a rotation $\pm \Delta$  in the  (X,Y)
 polarization of the string modes after passing the cosmic string.
 That is, there is no creation or excitation of modes after passing
 the scattering center [$\a^{A}_{n}$   and $(\a^{A}_{n})^{\dagger}$
   are not mixed]. In this elastic case, there is no
  particle transmutation (see sec.3).

	 Let us discuss now the inelastic scattering (case ii). The
 string evolution is described by the free equations except at the
 collision point  $(\s_{0} , \tau_{0})$   with the cosmic string. We take the
 origin at the cosmic string, thus we have
 $$
 X(\s_{0},\tau_{0})    =   0 \quad , \quad Y(\s_{0} ,\tau_{0})=0 \quad .
 \eqn\cont
 $$
 The values of  $\s_{0}$  and  $\tau_{0}$  indeed depend on the
 string state before the collision, that is, on the dynamical
 variables  $q^{X}_{<}, q^{Y}_{<}, p^{X}_{<} ,p^{Y}_{<} ,
 \a^{X}_{n<} , \tilde \a^{X}_{n<}, \a^{Y}_{n<}$ and  $\tilde
 \a^{Y}_{n<}$ [see below eq.(5.8)].

	 Since the deflection angles to the right and to the left of
 the cosmic string are different, we have now different matching
 conditions between the $<$ and the $>$ solutions
  for  $0 \leq \s  \leq \s_{0}$  and  for $\s_{0} \leq \s \leq 2\pi$,
 respectively.That is ,
 $$\eqalign{
 &X^{A}_{>}(\s,\tau_{0})  = X^{B}_{<}(\s,\tau_{0}){\cal R}_{B}^{A}( -\Delta)~~
  \cr &\partial_{\tau}X^{A}_{>}(\s,\tau_{0}) =
 \partial_{\tau}X^{B}_{<}(\s,\tau_{0}){\cal R}_{B}^{A}( -\Delta)~
 ,~~0 \leq \s  \leq \s_{0}~,A \equiv 1,2. \cr
  &X^{A}_{>}(\s,\tau_{0}) = X^{B}_{<}(\s,\tau_{0}){\cal R}_{B}^{A}( +\Delta)~~
 \cr
&\partial_{\tau}X^{A}_{>}(\s,\tau_{0})=\partial_{\tau}X^{B}_{<}(\s,\tau_{0})
 {\cal R}_{B}^{A}( +\Delta)~,~~\s_{0} \leq \s \leq 2\pi~, A \equiv 1,2.
 \cr }
 \eqn\bod
 $$
 Here  $X^1\equiv  X$  and  $X^2\equiv Y$.  This guarantees continuity
 of the string coordinates and its derivatives at $\tau  =  \tau_{0}$ .
 The string solutions $X^{A}_{<}(\s,\tau)$ and $X^{A}_{>}(\s,\tau)$   admit
 the usual expansion \smen . It should be also noticed that eqs.\bod\
  reproduce eqs.\rota\ with signs  +  or  -  when $\s_{0} = 0$ or
   $\s_{0}  = 2 \pi $ ,  respectively, as it must be.

 We recall that $\s_{0}$  and  $\tau_{0}$  depend on the initial data of the
 string through the constraint \cont\ as
 $$
 0 =  q^{A}_{<} \, +\, 2 \, p^{A}_{<} \, \a' \tau_{0} +
 i \sqrt{\a'} \sum_{n \neq 0}
   \{ \a^{A}_{n<} \exp[in(\s_{0} - \tau_{0})]
	 + \tilde  \a^{A}_{n<} \exp[-in(\s_{0} + \tau_{0} )] \}/n~,~A=1,2~~
 \eqn\conte
 $$
	 Imposing eqs.\bod\ at  $\tau = \tau_{0}$  yields linear relations
 between the operators $>$  and  $<$  appearing in the expansions
 \smen  . We find for the zero modes
 $$\eqalign{
 p_{>}^{A} =~ &p_{<}^{B} ~{\cal R}_{B}^{A}(\Delta) + {{\s_{0} +\tau_{0}}
 \o {2 \pi}} ~ p_{<}^{B} ~ {\cal L}_{B}^{A} + {1\o {4 \pi \a '}}~
 q_{<}^{B}~ {\cal L}_{B}^{A}  + \cr
 +  &{i \o {4 \pi \sqrt{\a'}}} \sum_{n \neq 0}  {{\exp[- in \tau_{0}]}
  \o n} \{\a^{B}_{n<} -
  \tilde  \a^{B}_{n<} ( 1 - 2 \exp[-in\s_{0}]) \}{\cal L}_{B}^{A} \cr
 q_{>}^{A} =~ &q_{<}^{B} ~{\cal R}_{B}^{A}(\Delta) + {{\s_{0} -\tau_{0}}
 \o {2 \pi}}~ p_{<}^{B}~{\cal L}_{B}^{A} - {{\a ' \tau_{0}^{2}} \o \pi}
p_{<}^{B}
  {\cal L}_{B}^{A}  + \cr
 +  &{\sqrt{\a'} \o {2 \pi }} \sum_{n \neq 0}  {{\exp[- in \tau_{0}]}
  \o {n^2}} \{\a^{B}_{n<}( \exp[in\s_{0}] - 1 - i n \tau_{0})+   \cr
 + & \tilde  \a^{B}_{n<} ( 1 -\exp[-in\s_{0}]+ i n \tau_{0}(1 - 2
 \exp[-in\s_{0}]) \} {\cal L}_{B}^{A} ~,~A=1,2 \cr}
 \eqn\mdcer
 $$
 where ${\cal L}_{B}^{A} \equiv {\cal R}_{B}^{A}( -\Delta) -
 {\cal R}_{B}^{A}( \Delta)$ . We get for the oscillator modes:
 $$\eqalign{
 \a^{A}_{n>}=&~\a^{B}_{n<}~{\cal R}_{B}^{A}(+\Delta)-{{i\sqrt{\a'}}\o{2\pi n}}
 \exp[in \tau_{0}]~[1 - \exp(-in\s_{0})]~[p_{<}^{B}(1 - i n \tau_{0})
 - {{in} \o {2 \a'}}q_{<}^{B}]~{\cal L}_{B}^{A} \cr
 &-{i \o {4 \pi}}\sum_{m \neq 0}  {{m+n} \o {m(m-n)}}\exp[ i(n-m) \tau_{0}]
  ~\{ \exp[-i(n-m)\s_{0}] - 1 \}~\a^{B}_{m<}~{\cal L}_{B}^{A} \cr
 &+{i \o {4 \pi}}\sum_{m \neq 0}(1/m)\exp[ i(n-m) \tau_{0}]~
  \{ \exp[-i(n+m)\s_{0}] - 1 \}~\tilde \a^{B}_{m<}~{\cal L}_{B}^{A}~, \cr
 \tilde \a^{A}_{n>}=&~\tilde\a^{B}_{n<}~{\cal R}_{B}^{A}(+\Delta)
 -{{i\sqrt{\a'}}\o{2\pi n}}
 \exp[in \tau_{0}]~[1 - \exp(in\s_{0})]~[p_{<}^{B}(1 + i n \tau_{0})
 + {{in} \o {2 \a'}}q_{<}^{B}]~{\cal L}_{B}^{A} \cr
 &-{i \o {4 \pi}}\sum_{m \neq 0}  {{m+n} \o {m(m-n)}}\exp[ i(n-m) \tau_{0}]
  ~\{ \exp[i(n-m)\s_{0}] - 1 \}~\tilde\a^{B}_{m<}~{\cal L}_{B}^{A} \cr
 &+{i \o {4 \pi}}\sum_{m \neq 0}(1/m)\exp[ i(n-m) \tau_{0}]
  ~\{ \exp[i(n+m)\s_{0}] - 1 \} ~\a^{B}_{m<}~{\cal L}_{B}^{A}~, \cr}
  \eqn\mdosc
 $$
 where the terms  $m = n$  in the sums are defined as their limiting values,
  $m \to n$.

	 The Bogoliubov transformations eqs.\mdosc\ are interpreted as usual
 (sec.3). In this inelastic case in which the string collides against
 the cosmic string,  besides the change of polarization, the modes
 become excited and then particle transmutation take place.
  As we have seen in sec.3 this phenomenon appears in general for
 strings propagating in curved (static as well as time
 dependent) spacetimes. Particle states get here transmuted at the classical
 (tree) level as a consequence of the interaction with the geometry.
 In the present case, this is a topological defect in the space-time.

	 It can be shown that the matching relations \mdcer -\mdosc\
 yield for the Virasoro generators $ L_{n}^{<}   =
L_{n}^{>}$  [\conico ] .
 In addition, the mass operators  $M_{<}^{2}$  and  $M_{>}^{2}$
 have identical spectra although they are not identical :  $M_{<}^{2}
 \neq M_{>}^{2}$.
 The critical dimension is the same as in Minkowski spacetime.

In the second reference under [\conico ] the superstring scattering by
a cosmic string is solved. That is, we exactly quantized NSR and
Green-Schwarz superstrings in the spacetime \conic . As for the purely
bosonic string, if the superstring touches the cosmic string, the
scattering is inelastic, since the internal superstring modes become
excited and mixing of particle- and antiparticle-modes takes place
leading to particle transmutation. We also generalized all these results
to the case when the cosmic string is spinning. The metric \conic\ when the
cosmic string possesses a spin $S$ takes the form
$$
 (ds)^2  = - (dX^{0} + {{GS} \o \a} d\Phi )^2  + (dR)^2 + R^2 ~(d\Phi )^2 +
 \sum_{i=3}^{D-1}(dX^{i})^2
 \eqn\spino
 $$
 Introducing the coordinate:
$$
\tilde X^0 = X^{0} + {{GS} \o \a} \Phi
$$
the metric becomes locally Minkowski and the points  $(\tilde X^0, R,
\Phi , X^{i})$ and  $(\tilde X^0 +2\pi G S , R, \Phi + 2\pi\a,X^{i})$
  are identified. As a consequence, besides the deficit angle $8 \pi G
\mu $ , the spacetime has a {\bf time-helical structure} in the coordinates
$(\tilde X^0 ,R, \Phi , X^{i})$ . That is, the spin introduces a shift
$2 \pi GS$ on $\tilde X^0$ , besides the rotation $2 \pi \a$ for two
identical points. For the bosonic string, the spatial coordinates are
given by the solution \smen\ as before. The new features when $S \neq
0$  appear in the coordinate $X^0$. We can choose here the light-cone
gauge
$$
X^{0} + {{GS} \o \a} \Phi + X^{1} = 2 \, \a ' p^U \tau
$$
For the fermionic coordinates, the propagation is not affected by the
spin of the cosmic string (fermion coordinates are invariant
under space-time translations).

 \chapter{Exact Integrability of strings in D-dimensional de Sitter
 spacetime}

 Let us consider the D-dimensional de Sitter space-time with metric given by
$$
ds^2=-dt_{0}^{2} ~+~ \exp[2Ht_{0}]~ \sum_{i=1}^{D-1} (dX^{i})^2 .
\eqn\medes
$$
 Here $t_{0}$ is the so called cosmic time. In terms of the conformal time
$\eta$,
$$
\eta \equiv-{{\exp[-H t_{0}]} \o H} \quad , -\infty < \eta \leq 0 ~ ,
\eqn\chid
$$
 the line element becomes
$$
ds^2 = {1 \o {H^{2} \eta^{2}}}[ - d\eta^2 ~+~ \sum_{i=1}^{D-1}
(dX^{i})^{2}\, ]  ~.
\eqn\mecon
$$
 The de Sitter spacetime can be considered as a D-dimensional hyperboloid
embedded in a D+1 dimensional flat Minkowski spacetime with coordinates
$(q^0,...,q^D)$ :
$$
ds^2 = {1\o {H^2} }[ - (dq^0)^2 ~+~ \sum_{i=1}^{D} (dq^i)^2 \, ]
\eqn\mehip
$$
 where
$$\eqalign{
q^0 =& ~\sinh{H t_0}~ + {{H^2} \o 2} \exp[H t_0]~ \sum_{i=1}^{D-1}
(X^i)^2 ~~, \cr
q^1 =& ~\cosh{H t_0}~ - {{H^2} \o 2} \exp[H t_0]~ \sum_{i=1}^{D-1}
(X^i)^2 ~~ , \cr
q^{i+1} =& ~H \exp[H t_0]~ X^{i} \quad , 1 \leq i \leq D-1, ~
-\infty < t_{0} , ~ X^{i} < + \infty .\cr}
\eqn\traco
$$
 The complete de Sitter manifold is the hyperboloid
$$
-(q^0)^2 + \sum_{i=1}^{D} (q^i)^{2} = 1.
\eqn\hiper
$$
 The coordinates  $(t_{0}, X^{i})$  and  $(\eta , X^{i})$  cover only
the half of the de Sitter manifold $q^0 + q^1 > 0$.

 We will consider a string propagating in this D-dimensional
space-time.The string equations of motion \conouno\ in the metric
\mehip\ take the form:
$$
\partial_{+-}q + (\partial_{+}q . \partial_{-}q)\;q = 0 \quad with \quad
q.q = 1 ,
\eqn\desmo
$$
 where . stands for the Lorentzian scalar product $a.b \equiv -a_0b_0 +
\sum_{i=1}^{D}a_ib_i ~ ,~x_\pm \equiv {1\o{2}}(\tau \pm \s)$ and
$\partial_{\pm}q = {{\partial q} \o {\partial x_{\pm}}}$.
The string constraints on the world sheet are
$$
T_{\pm\pm}= {\partial q \o {\partial x_\pm}}.
{\partial q \o {\partial x_\pm}} = 0           .                   \eqn\bincu
$$

 Eqs.\desmo\ describe a non compact O(D,1) non-linear sigma model in
two dimensions. In addition, the (two dimensional) energy-momentum
tensor is required to vanish by the constraints eqs.\bincu\ .
This system of non-linear partial differential equations can be
simplified by choosing an appropriate basis for the string coordinates
in the (D+1)-dimensional Minkowski space time $(q^0,...,q^D)$.
 The construction of this basis is analogous to the reduction of
 the O(N) non-linear sigma model [\pol ].
 We choose as a basis the vectors
$$
e_i = ( q, \partial_{+}q , \partial_{-}q , b_4 , ... , b_{D+1} )
\eqn\base
$$
 where the $b_i , 1 \leq i \leq D+1$ form an orthonormal set
$$
b_i . b_j = \delta _{ij} \quad and \quad b_i . q = b_i . q_\pm = 0
\eqn\ortg
$$
 We define
$$
\exp[\a(\s,\tau)] = -{\partial q \o { \partial x_-}}  . {\partial q \o
{ \partial x_+}}
                        \eqn\alfa
$$
 It is easy to show from eqs.\desmo\  and \bincu\  that
$$
q . \partial_{\pm}q =0 \quad \partial_{\pm}q . \partial_{\pm\pm}q = 0
\eqn\prods
$$
 In the basis \base\ , the second derivatives of $q$ expresses as [\desi
]
$$
\partial_{+-}q = q \exp \a \quad, \quad
\partial_{++}q = \partial_{+}\a \partial_{+}q +
\sum_{i=4}^{D+1} u_i b_i \quad, \quad
\partial_{--}q = \partial_{-}\a \partial_{-}q +
 \sum_{i=4}^{D+1} v_i b_i\quad .
\eqn\ders
$$
Here,
$$
u_i \equiv b_i . \partial_{++}q \quad , \quad v_i \equiv b_i .
 \partial_{--}q \quad .
$$
We find from eqs.\alfa\ - \ders\ [\desi ]
$$
 {{\partial^2 \a}\o{\partial \tau ^2 }} - {{\partial^2 \a}\o{\partial
\s ^2}} -\exp(\a) + \exp(-\a)~ \sum_{i=4}^{D+1} u_i v_i  = 0
  \eqn\gshg
$$
 This is the evolution equation for the function $\a(\s,\tau)$
determining the scalar product $q_+ . q_- $, for all D. This is a
generalization of the Sinh-Gordon equation. It remains to find now the
evolution equations for the fields $u_i$ and $v_i$ . In order to find such
equations we express the derivatives of the basis
vectors eq.\base\ in terms of the basis itself (Gauss-Weingarten equations):
$$
\partial_+ e_i = A_{ij}(\s,\tau) e_j \quad , \quad
\partial_- e_i = B_{ij}(\s,\tau) e_j
\eqn\gaco
$$
 The compatibility condition for eqs.\gaco\ expresses as
$$
\partial_- A - \partial_+ B + [ A , B ] = 0
\eqn\lax
$$
We want to stress that  $\exp[\a(\s,\tau)]$  has a clear physical
interpretation. The invariant interval between two points on the
string computed with the spacetime metric \mehip\ is given by
$$
ds^2 ={1\o{H^2}}dq.dq = {1\o{2H^2}} \; \exp[\a(\s,\tau)]\;(d\s^2-d\tau^2)
                                                        \eqn\intcuer
$$
We see that the factor $\exp[\a(\s,\tau)]$ determines the string size.

  Let us start studying the case D = 2. In this case, a complete
basis is formed by
$$
e_i = ( q , \partial_{+}q , \partial_{-}q )~.
$$
 Therefore, eqs.\ders\ and \gshg\ become :
$$
\partial_{++}q =\partial_{+} \a \partial_{+}q \quad , \quad
\partial_{--}q = \partial_{-}\a \partial_{-}q
$$
 and
$$
{{\partial^2 \a}\o{\partial \tau ^2 }} - {{\partial^2 \a}\o{\partial
\s ^2}} -\exp\a  = 0 ~.
\eqn\liou
$$
 This is the Liouville equation whose general solution is given by
$$
\a(\s,\tau) = \log{{2 f'( x_+) g'( x_-)} \o {[ f( x_+) + g( x_-)]^2}}~,
\eqn\soge
$$
 where $f$ and $g$ are arbitrary functions of the indicated variables.

 The D = 2 case can be solved directly from the equations of motion
\desmo\  and the constraints eqs.\bincu\  in the coordinates $(u,v)$
defined as follows:
$$\eqalign{
q_0 = \sinh u~,~q_1 =& \cosh u \cos v ~,~ q_2 = \cosh u \sin v \cr
&-\infty < u < +\infty~,~ 0 \leq v < 2 \pi.\cr}
\eqn\uv
$$
 The constraints eqs.\bincu\ take the form
$$
(\partial_{\pm}q )^2 = (\partial_{\pm}v \cosh u )^2 - (\partial_{\pm}u)^2 = 0
\eqn\vndos
$$
 Therefore, we have
$$
\partial_{+}v \cosh u = \pm \partial_{+}u \quad , \quad
\partial_{-}v \cosh u = \pm \epsilon ~\partial_{-}u \quad ,
\eqn\visol
$$
 where $\epsilon ^2 = 1$ . In addition,
$$
\partial_{+}q.\partial_{-}q = (\cosh u )^2 \partial_{+}v
 \partial_{-}v - \partial_{+}u \partial_{-}u = (\epsilon - 1)
\partial_{+}u  \partial_{-}u
\eqn\esca
$$
 The general solution of eq.\visol\ is given by
$$
v = \pm 2 \arctan[\exp(u)] + G(x_-) \quad , \quad
v = \pm 2 \epsilon \arctan[\exp(u)] + F(x_+) \quad ,
\eqn\solex
$$
where $F$ and $G$ are arbitrary functions of the indicated variables.
We have here two different cases, depending on whether
A : $\epsilon = -1$ or B : $\epsilon = +1$

 {\bf Case A : $\epsilon = -1$}

 Here, we find
$$
v = {1 \o 2}[ F(x_+) + G(x_-)] ~,~ u= \log[\pm \tan({{F-G} \o 4})]
\eqn\sola
$$
 This corresponds to the previous solution eq.\soge\ with
$$
f(x_+) = -\exp[iF(x_+)] \quad g(x_-) = \exp[iG(x_-)] ~~~.
$$
 Using the world sheet conformal invariance, we can always choose the gauge
where
$$\eqalign{
&v = \s ~,~ 0 < \s \leq 2\pi~,~~ u=\log(\tan {\tau \o 2})~~,~~ 0 < \tau
\leq \pi,\cr &\sinh u = - \cot{\tau} ~,~ -\infty < u < +\infty .\cr}
\eqn\solaa
$$
 This describes a string winded around the de Sitter universe and
evolving with it. A half of string evolution $\pi /2 < \tau < \pi$
corresponds to the expansion time $0 \leq u < \infty$  of the de Sitter
universe. Similarly, for the first half $0 < \tau < \pi/2$, which corresponds
to the contraction phase $-\infty  < u \leq 0$. (see fig.3).
 Eq.\solaa\  describes a string winded once around de Sitter space
(here a circle). More generally, we may have
$$
v = n \s  ~,~ 0 < \s \leq 2\pi~,~~ u=\log(\tan {{n\tau} \o 2})~,
\eqn\soln
$$
where $n$ is an integer number. This solution describes a string winded
$n$ times around the de Sitter space.

 Let us consider the invariant interval \mehip\ between two points on the
string using coordinates $(u,v)$,
$$
ds^2 = {1 \o {H^2}} ~[ - du^2 + \cosh ^2 u ~dv^2 ]~.
$$
 For the solution eq.\soln\ , we have
$$
ds^2 = ({n \o {H \sin n\tau }})^2 ( d\s^2 - d\tau^2 )~.
$$
 In the asymptotic regions $\tau \to 0^+$ and $\tau \to \pi^- /n$,
the conformal factor blows up. The proper length of the string
stretches infinitely as
$$\eqalign{
\Delta s =& {1 \o {H \tau}}~\Delta \s ~~for~~\tau \to 0^+ \cr
\Delta s =& {1 \o {H (\pi/n-\tau)}}~\Delta \s ~~for~~\tau \to \pi^- /n~.
\cr}
\eqn\estal
$$
 This is analogous to the unstable behaviour found in D-dimensional
inflationary backgrounds (see below and refs.[\vene ]), as well as
for strings falling into space-time singularities (see sec.4 and
refs.[\ondpl ]).

 {\bf Case B : $\epsilon = +1$}

  In this case,
$$
\partial_{+}q . \partial_{-}q = 0 \quad \partial_{+-}q = 0
\eqn\tres
$$
 and therefore, the parametrization in terms of the field
$\a(\s,\tau)$  breaks down.
 Eqs.\solex\ yields $F = G = constant \equiv C$ . Then,
$$
u = \log[ \pm \tan({{v-C} \o 2})]
\eqn\solb
$$
 Now, we consider the equation of motion \tres\ to find the dependence
on $x_+$ and $x_-$. Using also eqs.\uv\ and \solb\ , we find
$$\eqalign{
q_0 = & \mp \cot(v-C) ~,~q_1 = \pm \cot(v-C) ~ \cos C \mp \sin C ~, \cr
q_2 = & \pm \cot(v-C) \sin C \pm \cos C \cr}
\eqn\qub
$$
 Therefore, we obtain
$$
v = C - \pi / 2 + \arctan[ R(x_+) + S(x_-)] ~,~ \cosh u =
\sqrt{1 + [ R(x_+) + S( x_-)]^2 }
\eqn\qubb
$$
 where $R( x_+)$ and $S( x_-)$ are arbitrary functions of the indicated
variables.

 Using the world sheet conformal invariance, we can always choose
$$
v = C - \pi /2 \pm \arctan \tau \quad \cosh u = \sqrt{ 1 + \tau ^2}
\eqn\solbb
$$
 Therefore, the string solution eqs.\qub\ yields
$$
q_0 = \tau ~,~q_1 = -\tau \cos C - \sin C~,~ q_2 =- \tau \sin C
+ \cos C
\eqn\bfin
$$
 This solution actually describes a particle trajectory since it was
possible to gauge out the parameter  $\s$ . Eq.\bfin\ describes a
geodesic in the two dimensional de Sitter space-time, that is the
trajectory of a massless particle. Since transverse dimensions are
absent, only massless states appear in this two-dimensional case.
The solution eq.\solbb\ or \bfin\ is a particular case of the center
of mass solution described in ref.[\nos ] when D = 2 and $m = 0$.  When $\tau$
goes from $-\infty$ to $+\infty$ , the light rays go from  $q^0=
-\infty$  to  $q^0 = + \infty$ .  At the same time the angle $v$ varies
 through an intervale of $\pi$ :  $v ( -\infty) = v(+\infty) \pm \pi $
. In eq.\solbb\ , the signs $\pm$ correspond to a motion in the
positive  or negative direction of the de Sitter spatial circle
(see fig.4). It should be noticed that travelling from
$q^0 = \tau = -\infty$  to  $+\infty$ , the particle goes over
half of the de Sitter circle. The solutions described in the
A and B cases, contain {\bf all the string solutions} in the two-dimensional
de Sitter space-time.

Let us now consider strings in the 2+1-dimensional de Sitter spacetime.
There, we have a four dimensional embedding Minkowski
space-time where the antisymmetric Levi-Civita tensor allows us to
construct a vector $b \equiv b_4$ orthogonal to the vectors
$q, \partial_{+}q$ and $\partial_{-}q$ , namely
$$
b_a \equiv \exp(-\a)~ \epsilon_{abcd}\, q_b \, (\partial_{+}q)_c \,
(\partial_{-}q)_d
\eqn\vecb
$$
 The vectors $(q, \partial_{+}q, \partial_{-}q, b)$ form a basis. In addition,
$b .b = 1 $ .
 Here, the compatibility condition eq.\lax\  yields
$$\eqalign{
u \equiv u_4 =  u(x_+) ~~,~~ v \equiv & v_4 = v(x_-) \cr
{{\partial^2{\a}}\o{{\partial{x_-}\partial{x_+}}}}-\exp\a +}
\eqn\uvpm
$$
Upon a conformal transformation $x_{\pm} \to \Phi(x_{\pm}) ,\a \to
\a + \log[ \Phi_{+}' \Phi_{-}']$ with $ (\Phi_{+}')^2 = u(x_+)$
and $(\Phi_{-}')^2 = v(x_-)$, eq.\uvpm\ takes the Sinh-Gordon form
$$
 {{\partial^2 \a}\o{\partial \tau ^2 }} - {{\partial^2 \a}\o{\partial
\s ^2}} -\exp\a + \exp-\a  = 0
                                                \eqn\shG
$$
Notice that for closed strings $q(\s,\tau)$ and hence $\a(\s,\tau)$
are periodic functions of $\s$ with period $2\pi$.
Therefore, to find string solutions in de Sitter spacetime we can start
from a periodic solution of eq.\shG\ , and insert it on the field
equations \desmo :
$$\eqalign{
[{{\partial^2 }\o{\partial \tau ^ 2}} -
{{\partial^2 }\o{\partial \s ^2}}
 - \exp\a(\sigma,\tau)]q(\sigma,\tau)=0 \cr}
\eqn\kgg
$$
Once this {\bf linear} equation in $q(\sigma,\tau)$ is solved, it
remains to impose the constraints \bincu\ and eq.\alfa.

The energy density for the sinh-Gordon model \shG\ reads here
$$
\H = {1\o2}[({{\partial \alpha}\o{\partial \tau}})^2+({{\partial
\alpha}\o{\partial \s}})^2] - 2\cosh\a(\sigma,\tau)
                                        \eqn\energia
$$
That means a potential {\bf unbounded from below}
$$
V_{eff} = - 2 \cosh\a   \quad  ,
                                        \eqn\potencial
$$
with absolute minima at $\a = +\infty$ and at $\a = -\infty$.
As the time $\tau$ evolves, $\a(\s,\tau)$ will generically approach
these infinite minima. The first minimum corresponds to an infinitely large
string whereas the second describes a collapsed situation. That means
that strings in de Sitter spacetime will generically tend either to
{\bf inflate} at the same rate as the universe (when $\a\to+\infty$) or to
{\bf collapse} to a point (when $\a\to-\infty$).
As we shall see below these general trends are confirmed by the
explicit string solutions.

Let us start by studying solutions where  $\a=\a(\tau)$. Then the
energy is
$$
{1\o2}{\alpha^{\prime}}^2  - 2 \cosh\a = E = constant \geq -2
                                \eqn\enerII
$$
$\a(\tau)$ describes the position of a non-relativistic particle with
unit mass rolling down the effective potential $V_{eff}=-2\cosh\a$ . A
particularly interesting situation is the critical case $E = -2$ when
one starts to roll down from the maximun of $V_{eff}$. That is, the
initial speed is zero and the 'time' $\tau$ to reach either minimun
($\a=+\infty \;  or \;  -\infty$) is infinity. The corresponding solutions are
$$
\a_-(\tau)= \log[\coth^2 {\frac{\tau}{\sqrt{2}}}]\quad   and \quad
\a_+(\tau)= \log[\tanh^2 {\frac{\tau}{\sqrt{2}}}]
                                \eqn\solualfa
$$
$\a_-(\tau)$ starts at $\a=0$ for $\tau=-\infty$ and rolls down to the
{\bf right} reaching $\a=+\infty$ for $\tau\to 0^-$.
The solution $\a_+(\tau)$ also starts at $\a = 0$ for $\tau = -\infty$
but rolls down to the {\bf left} reaching $\a=-\infty$ for $\tau \to
0^-$. Notice that $\a_+(\tau) = -\a_-(\tau)$. In addition we have the
trivial (but exact) solution $\a^o(\tau)\equiv0$.

Now that the function $\a(\tau)$ is known, we proceed to solve
eq.\kgg\   for $q(\s,\tau)$ with the constraints \bincu\ and \alfa\ .
 Since $q_0$ is a time-like coordinate, we shall assume $q_0 =
q_0(\tau)$.
A natural ansatz is then
$$
q=(q_0(\tau),q_1(\tau), f(\tau) \cos\s, f(\tau) \sin\s )
                                                \eqn\ansatzI
$$
 Then, eqs.\desmo\ , \bincu\ and \alfa\  require
$$
q_{0}(\tau)^{2} = q_{1}(\tau)^2 + f(\tau)^2
                                \eqn\ufaI
$$
$$
[{{dq_0}\o{d\tau}}]^2=[{{dq_1}\o{d\tau}}]^2 + [{{df}\o{d\tau}}]^2 + f^2
                                \eqn\ufaII
$$
$$
e^{\a(\tau)}=[{{dq_0}\o{d\tau}}]^2-[{{dq_1}\o{d\tau}}]^2 - [{{df}\o{d\tau}}]^2
+ f^2
                                \eqn\ufaIII
$$
and
$$\eqalign{
{{d^2}\o{d^2\tau}}q_0 - e^{\a(\tau)}q_0(\tau)=0 \cr
{{d^2}\o{d^2\tau}}q_1 - e^{\a(\tau)}q_1(\tau)=0 \cr
{{d^2}\o{d^2\tau}}f(\tau) + f(\tau)-e^{\a(\tau)}f(\tau)=0 \cr}
                                \eqn\ansat
$$
In addition, it seems reasonable to choose the time coordinate
$q_0(\tau)$ to be an odd function of $\tau$. Remarkably enough,
eqs.\ufaI\ - \ansat\ admit consistent solutions for $\a(\tau) =
\a_+(\tau)$, $\a(\tau) = \a_-(\tau)$ and $\a(\tau) = 0$. We find for
$\a(\tau) = \a^o(\tau) = 0$
$$
q^o(\s,\tau)= {1\o{\sqrt{2}}}(\sinh\tau,\cosh\tau,\cos\s,\sin\s)
                                        \eqn\solI
$$
For $\a(\tau) = \a_-(\tau)$, we have
$$\eqalign{
q_-(\s,\tau) =
(\sinh\tau -{1\o{\sqrt{2}}}\cosh\tau \,
\coth[{1\o{\sqrt{2}}}\tau],\quad
\cosh\tau -{1\o{\sqrt{2}}}\sinh\tau \,
\coth[{1\o{\sqrt{2}}}\tau],\cr
{1\o{\sqrt{2}}}\cos\s \,
\coth[{1\o{\sqrt{2}}}\tau],\quad {1\o{\sqrt{2}}}\sin\s \,
\coth[{1\o{\sqrt{2}}}\tau]) , \qquad  \qquad \cr}
                                        \eqn\solII
$$
And for $\a(\tau) = \a_+(\tau)$ we find
$$\eqalign{
q_+(\s,\tau)=
(\sinh\tau -{1\o{\sqrt{2}}}\cosh\tau
\tanh[{1\o{\sqrt{2}}}\tau],\quad
\cosh\tau -{1\o{\sqrt{2}}}\sinh\tau
\tanh[{1\o{\sqrt{2}}}\tau],\cr
{1\o{\sqrt{2}}}\cos\s
\tanh[{1\o{\sqrt{2}}}\tau], \quad {1\o{\sqrt{2}}}\sin\s
\tanh[{1\o{\sqrt{2}}}\tau]) \qquad \qquad . \cr}
                                          \eqn\solIII
$$
 These string solutions are given for a fixed de Sitter frame.
Applying the de Sitter group to them yields a multi-parameter family of
solutions. As it is clear, we can study them in the frame
corresponding to eqs.\solI\ - \solIII\ without loss of generality.
Let us now discuss the physical interpretation of these solutions.

We recall that for a given time  $ q_0 = q_0(\tau) $, the de Sitter
space is a sphere $ S^2$ with radius $R(\tau) = {1\o{H}}\sqrt{1+q_0(\tau)^2}$.
$q^o(\s,\tau)$ [eq.\solI ], describes a string of constant size in a
de Sitter
universe that inflates for $\tau\to\infty$ since for this solution
$  R(\tau) = {1\o{H}}\sqrt{1+{\sinh^2 \tau \o{2}}} $.
This solution is probably unstable under small perturbations.

The solution $q_-(\s,\tau)$ is more interesting. One should
distinguish four domains in it:

(i) $-\infty < \tau < -\tau_0 $, (ii) $-\tau_0 < \tau < 0$, (iii) $ 0 <
\tau < \tau_0$ , (iv)  $\tau_0 < \tau < +\infty$ ,

where $ q(\tau_0) = q(-\tau_0) = 0$ .

{}From eq.\solII\ we find $\tau_0 = 1.489...$ .
In the intervals (i) and (iii) $R(\tau)$ decreases (the universe
contracts), whereas for (ii) and (iv)  $R(\tau)$  grows, (the universe
expands).
The string size is given here by
$$
S_-(\tau) = {1\o{\sqrt{2}H}}\coth[{1\o{\sqrt{2}}}|\tau|]
                                        \eqn\talla
$$
That is, the string size increases for $\tau < 0$ and decreases for $
\tau > 0$ with a singular behaviour ${1\o{|\tau}|}$ for $\tau \to 0$.
We see that the string size grows monotonically in intervals (i) and
(ii), this growing becoming explosive for $\tau \to 0$ when the size
of the de Sitter space diverges. Actually, the string grows there at
the same rate as the whole space $(\simeq{1\o{|\tau|}})$ .

 For large $|\tau|$ the de Sitter space is also very large with
$$
R_-(\tau)\quad \buildrel{|\tau| \to \infty}\over= \quad
{{\sqrt{2}-1}\o{2\sqrt{2}H}} \;  e^{|\tau |}\; [1 + O(e^{-|\tau|})]
                                        \eqn\erre
$$
whereas the string size tends to a constant
$$
S_-(\tau) \quad  \buildrel{|\tau| \to \infty} \over = \quad {1\o{ \sqrt{2}H}} +
O(e^{-\sqrt{2}|\tau|})
                                        \eqn\ese
$$
The behaviour for small $|\tau|$ confirms the asymptotic results found in
refs.[\nos\ - \desi\ - \vene ].

It is interesting to study this string solution in another set of de Sitter
coordinates. The cosmic time $t_0$ and the conformal time $\eta$
 [eq.\chid ] take for $q_-(\s,\tau)$ the form:
$$
e^{Ht_0} = - {1\o{H\eta}}= [1 - {1\o{\sqrt{2}}}
\coth({1\o{\sqrt{2}}}\tau)] \; e^{\tau} ~,~
\rho = {{e^{-\tau}}\o{[1 - \sqrt{2}
\tanh({1\o{\sqrt{2}}}\tau)]}} \quad
   (\tau < 0).
  \eqn\coorden
$$
where
$$ \rho \equiv {\sqrt{(q_2)^2+(q_3)^2}\o{H(q_0 + q_1)}}.
                                        \eqn\ro
$$
Therefore, for $\tau \to 0$ (large universe and inflating string),
$$
\eta = {{\tau}\o{\pi}}\to 0 ,   \quad
\rho = 1 + O(\tau^2) \quad ,\quad
t_0 = -{1\o{H}}\log|\tau| + O(\tau) \to \infty
                        \eqn\otra
$$
whereas for $\tau \to -\infty$ (large universe but fixed string size),
$$
t_{0} = {{\tau}\o{\pi}} + {1\o{H}}\log(1+\sqrt{2}) \to -\infty ~,~
\eta = -{{e^{-\tau}}\o{H(1+{1\o{\sqrt{2}}})}} \to -\infty ~,~
\rho = {{e^{-\tau}}\o{(1 + {1\o{\sqrt{2}}})H}} \to \infty ~.
\eqn\otramas
$$
We see that $\tau$ interpolates between the cosmic and the conformal
time.
Notice that this confirms the asymptotic behavior \otra\ discussed in previous
works[\nos , \desi , \vene ].

Let us now discuss the solution $q_+(\s,\tau)$ [eq.\solIII ]. There
are here two phases:

(i) $\tau < 0$ : contraction phase, $R(\tau)$ decreases,

(ii) $\tau > 0$ : expansion phase, $R(\tau)$ grows.

The string size is here
$$
S_+(\tau) = {1\o{\sqrt{2}H}}\tanh[{1\o{\sqrt{2}}}|\tau|]
                                        \eqn\tallaII
$$
Therefore, the string contracts from a fixed size ${1\o{\sqrt{2} H}}$
at $\tau = -\infty$ during (i) till the colapse at $\tau = 0$. At this
point the de Sitter space has a minimun size  ${1\o{H}}$ . For $\tau > 0$
, the string size grows from zero till the fixed
value ${1\o{\sqrt{2}H}}$ , while the de Sitter space radius tends to
infinity as
$$
R_+(\tau) \buildrel{\tau\to\infty}\over= \; (\sqrt{2} - 1
){1\o{2\sqrt{2}H}}e^\tau
                        \eqn\rgrande
$$
This behaviour is quite different from $q_-(\s,\tau)$ and was not
noticed before.
Additional solutions follow by replacing
$$  \s \to n\s ,\quad   \tau \to n \tau  ,\quad  n \e Z
        \eqn\rulo
$$
in eqs.\solI\ - \solIII\  . In these solutions the string is winded $n$
times around the $q_1$ axis.

In addition, eq.\enerII\ leads to elliptic solutions for $E > -2$. However,
as shown in [\desi ], the string constraints select periodic solutions
of the sinh-Gordon equation associated to the lower boundary of the
allowed zone, therefore excluding elliptic solutions. More precisely,
we found  (real) elliptic solutions of the sigma model field equations
 \desmo\ for $0 \geq E > -2$ {\bf but not} solutions of the string equations
since $ T_{\pm\pm} $ turned to be a non-vanishing
constant for $E > -2$ [\desi ].

As last topic, let us consider strings propagating in the 3 + 1
dimensional de Sitter spacetime.
Here the reduced model \gshg\ contains in principle four fields ($u_4
, u_5 , v_4$ and $v_5$) besides $\a(\s,\tau)$ . Using conformal
invariance and some O(3,1) invariance, we eliminate all these fields
except one : $\beta (\s,\tau)$. $\beta (\s,\tau)$ is defined through:
$$
U(x_+) V(x_-)\cos\beta (\s,\tau) = u_4 v_4 + u_5 v_5
$$
where $U(x_+) = V(x_-)=1$ in an appropiate conformal frame [\desi ].
The equations of motion for $\a(\s,\tau)$ and  $\beta (\s,\tau)$
can be derived from the (reduced) Lagrangian
$$
\L = {1 \o 2}[(\partial_{\tau} \a)^2 -(\partial_{\s} \a)^2] +
 {1 \o 2}[(\partial_{\tau} \beta)^2 -(\partial_{\s} \beta)^2] -
V(\a,\beta)~~,~~V(\a,\beta)= -e^{\a} -e^{-\a}\cos{\beta}.
\eqn\toda
$$
 This potential can be related to the $B_2$ Toda field theory upon
changing $\beta \to i \beta$ . It is a hyperbolic Toda model connected
with the O(3,1) group instead of the O(4) group.

The potential \toda\ is unbounded from below (as \energia -\potencial\
 in 2+1 dimensional de Sitter) and indicates that the string
time evolution will tend to the absolute minima at  $\a = +\infty$
or  at $\a = -\infty$ with $|\beta| < \pi/2$ [\desi ]. In ref.\desi\ we
derive the string behaviour near such points.
We see from eq.\gshg\ that   $\a = +\infty$ and  $\a = -\infty$
are present as strongly attractive points for any dimension D. In
other words, strings in D-dimensional de Sitter spacetime tend
generically to inflate an sometimes to collapse as we have seen
explicitly in concrete 2+1 dimensional solutions.

\chapter{Concluding Remarks}

Inexpected and deep similarities appear between the string behaviour
near spacetime singularities  in singular plane waves (Sec.4) and
for expanding  (non-singular) universes (sec. 6).
The $U = 2 \a ' p \tau \to 0$ string behaviour
in the $\b = 2$ singular plane wave case [eqs.\bestia ] is similar to
the inflating string
solution $q_-(\s ,\tau)$ in de Sitter spacetime [eqs.\solII\ and
\otra ] for $\tau \to
0^- $.  $U$ plays in the first case a similar r\^ole than the conformal
time $\eta$ in the de Sitter case. Both $U$ and $\eta$ are proportional
to $\tau$ in these regimes. Moreover,
 the string blows up in both cases with its size growing as a power
of $ 1/\tau $ for $\tau \to 0$.  We have
$|\tau^{1-\sqrt(1+4\bar\a)}|$ and $|\tau|^{-1}$, for singular plane
waves ($\b =2$) and de Sitter respectively. In addition, for FRW
universes with conformal factor proportional to $(\eta)^{-\g} ,( \g >
0) $ ,  the string
size blows up as $|\tau|^{-\g/2}$ for $ \tau \to 0 $
[\vene ]. This entails a continuous
exponent, as for singular plane waves.

The fact that similar features  appear in such different
geometries strongly indicates
that the string stretching phenomenon is a generic property for strings on
strong gravitational fields.

	The aim of these lectures is to present the basic notions about strings in
curved spacetime. This is the begining of a vast and relevant domain. We think
that it will be helpful and necessary for the quantum understanding of gravity.
\refout

\chapter{Figure Captions}

\item{Fig.1.}

Escape string directions. Horizontal lines : $\a_1 > 0$ , vertical
lines : $\a_1 < 0$ .

\item{Fig.2.}

Particle trajectories in a conical spacetime with defect angle $ 2\pi(1-
\a )$.
\item{Fig.3.}

 The one-sheet hyperboloid represents the 1+1-dimensional de Sitter
spacetime embedded in a three-dimensional space. The closed circle
represents a string solution A [eq.\solaa ] at a given time.
\item{Fig.4.}

 Same as in fig.3 but now the string solution B [eq.\bfin ] is drawn.
This is in fact a geodesic.

\bye